\documentclass[11pt]{article}
\usepackage{amsmath}
\usepackage{amsfonts}
\usepackage{bbm}
\usepackage{bm}
\usepackage{graphicx}
\usepackage{enumerate}
\usepackage[semicolon]{natbib}
\usepackage{url}

\usepackage{caption}
\usepackage{setspace}
\usepackage{wrapfig}
\usepackage{appendix}
\usepackage{float}

\newcommand{\blind}{1}

\usepackage[export]{adjustbox}
\usepackage{wrapfig}
\usepackage{framed}
\usepackage{xcolor}
\usepackage{authblk}
\begin{document}

\def\spacingset#1{\renewcommand{\baselinestretch}%
{#1}\small\normalsize} \spacingset{1}


\if1\blind
{
  \title{\bf Joint species distribution modeling of abundance data through latent variable barcodes}
    \author[1]{Braden Scherting}
    \author[2]{Otso Ovaskainen}
    \author[1]{David B. Dunson}
    \affil[1]{Department of Statistical Science, Duke University}
    \affil[2]{Department of Biological and Environmental Science, University of Jyv\"askyl\"a; P.O. Box 35, 40014 Jyv\"askyl\"a, Finland.}
  \maketitle
} \fi

\if0\blind
{
  \bigskip
  \bigskip
  \bigskip
  \begin{center}
    {\LARGE\bf 
Inferring latent structure in ecological communities via barcodes     
    }
\end{center}
  \medskip
} \fi

\bigskip
\begin{abstract}
Accelerating global biodiversity loss has highlighted the role of complex relationships and shared patterns among species in determining their responses to environmental changes. The structure of an ecological community, represented by patterns of dependence among constituent species, signals its robustness more than individual species distributions. We focus on obtaining community-level insights based on underlying patterns in abundances of bird species in Finland. We propose \texttt{barcode}, a modeling framework to infer latent binary and continuous features of samples and species, expanding the class of concurrent ordinations. This approach introduces covariates and spatial autocorrelation hierarchically to facilitate ecological interpretations of the learned features. By analyzing 132 bird species counts, we infer the dominant environmental drivers of the community, species clusters and regions of common profile. Three of the learned drivers correspond to distinct climactic regions with different dominant forest types. Three further drivers are spatially heterogeneous and signal urban, agricultural, and wetland areas, respectively. 
\end{abstract}

\noindent%
\vfill

\spacingset{1.9}
\section{Introduction} \label{sec:intro} 

Understanding the structure of ecological communities is central to studying biodiversity and the processes that sustain it. Ecological communities are assemblages of organisms that occupy a shared environment and interact in potentially complex ways. Communities arise through the interplay of environmental filtering, species interactions, and stochastic dynamics, all of which depend on species- and habitat-specific traits \citep{kraft2015community}. Birds, in particular, often have an outsized impact on the ecosystems and communities they occupy \citep{leito2016black} and exhibit strong interspecific dependence \citep{lovette2006simultaneous, skorka2014invasive}, which can moderate or amplify responses to environmental change \citep{ahola2007climate, engelhardt2020ignoring}. Identifying community-wide environmental drivers and shared species responses is essential for effective biomonitoring and conservation.
Disentangling the drivers of community composition requires statistical tools that can accommodate the joint responses of multiple species to complex environmental and biological gradients. Species often exhibit correlated distributions due to shared ecological preferences, interactions, or unmeasured environmental factors \citep{tilman1982resource, chase2009ecological}. These dependencies confound single-species analyses, motivating the use of multivariate techniques that can identify common patterns in community data \citep{warton2015so}. 

A principal challenge in ecological community modeling is deciphering patterns of relative abundance among dozens or hundreds of species. Ordination is a class of multivariate statistical procedures designed to address this challenge \citep{legendre2012numerical}. Ordination methods project observed community data onto a lower-dimensional latent space, and variation in this latent space represents the environmental gradients that drive community structure. Ordination scores are often used in exploratory analyses to visualize pairwise relationships or cluster species. These tasks can be performed without directly interpreting the latent dimensions in terms of known habitat, climate, or evolutionary information. Alternatively, ordinations axes can be constrained to follow functions of environmental covariates. Methods that constrain the ordination axes in this way are termed constrained ordination, and methods that do not are termed unconstrained ordination \citep{ter1988theory}. Comparison of results between these approaches can give an indication of the extent to which measured environmental variables capture dominant underlying environmental gradients. 
Classical ordination methods, such as principal coordinate analysis \citep{gower1966some} and non-metric multidimensional scaling \citep{kruskal1978multidimensional}, are algorithmic or optimization-based. Model-based ordinations, which invoke a stochastic, often distributional relationship between an ordination and multivariate data, have grown in popularity. Model-based ordination can more flexibly represent abundance data, while enabling model checking and uncertainty quantification \citep{hoegh2020evaluating}. Popular approaches to model-based ordination include reduced rank regression \citep{davies1982procedures, yee2003reduced} and generalized linear latent variable models \citep{hui2015model}. We employ a constrained model-based ordination to derive and interpret dominant environmental gradients among Finnish birds.

GLLVMs \citep{skrondal2004generalized} are popular in ecology for both ordination and joint species distribution modeling (JSDM)  \citep{niku2019gllvm, ovaskainen2017make}.  \cite{hui2017model} and \cite{stratton2024clustering} use Gaussian mixtures for site scores to infer clusters in an unconstrained ordination approach. To unite advantages of constrained and unconstrained ordination, \cite{van2023concurrent} model site scores as stochastic functions of covariates within a GLLVM, an approach they term concurrent ordination. We build on these developments to simultaneously group samples and species, and learn interpretable environmental gradients. Recent extensions of latent class models using highly structured and identifiable binary latent variables \citep{gu2023bayesian, zhou2024bayesian} showcase the potential of discrete variables to flexibly characterize joint distributions of categorical data. In addition to being intrinsically interpretable, binary latent variables also induce natural clustering without the usual mixture model specification. We integrate these developments into an additive latent variable model.

Modeling of species abundance or relative abundance on the logarithmic scale has a rich history, motivated by the multiplicative nature of ecological processes such as reproduction and resource acquisition \citep{preston1948commonness, watterson1974models}. Log-linear Poisson and negative binomial models remain standard in community ecology \citep{ hui2015model}. However, additive formulations (i.e., identity link) offer both modeling and inferential advantages. Additive models can better accommodate species with both low or zero counts and occasional large counts, which often violate distributional assumptions or lead to numerical instability in log-scale models. Additive non-negative factors are also straightforward to interpret, unencumbered by nonlinear transformations. 
Our approach, dubbed \texttt{barcode} (\textbf{b}inary \textbf{a}nd \textbf{r}eal \textbf{co}unt \textbf{de}composition), utilizes additive latent variables to learn the primary drivers of relative abundance within a large bird community. Binary latent variables and latent covariate effects aid in the interpretation. 

\section{Dataset and Scientific Questions}\label{sec:data}
National biodiversity monitoring programs are common in many regions and are important tools for informed conservation. Bird monitoring is popular because birds are relatively easy to detect and identify, and they occupy large geographic regions and ecological niches. Fennoscandian bird monitoring surveys have proven valuable in studying declines in mountain species \citep{lehikoinen2014common} and trends in wader species \citep{lindstrom2015large, lindstrom2019population}. To better understand broad trends and borrow information among species groups, we use line transect counts of 132 species from the Finnish national bird monitoring program documented by \cite{piirainen2023species}. 
Line transect count surveys (sampling units) were conducted by volunteers beginning in 1978. The set of study species was expanded in 2006 to include waterbirds. We limit our focus to surveys conducted between 2006 and 2016. This 11 year period includes 2826 surveys  at 555 sites; sites were visited between 1 and 11 times each. An average of 260 (1–759) individuals and 38 (1–81) species are recorded in each sample. Because survey effort is limited and detection imperfect, the recorded count of each species is assumed to reflect that species' relative abundance in the community. This assumption is limiting if detectability is substantially different between species, although our focus on inferences at the community level rather than species level lends robustness. Therefore, ``abundance'' refers not to absolute abundance but to this sample abundance. 

On average, each survey recorded 40 willow warblers (\textit{Phylloscopus trochilus}; most abundant) and 0.03 ospreys (\textit{Pandion haliaetus}; least abundant). Approximately $72\%$ of the records are zero, and the largest single count is 189 (black-headed gull, \textit{Larus ridibundus}). Sample- and site-specific auxiliary information is also available. 
The starting points of the line transects inform the geographic abundance gradients across Finland, and the lengths of the line transects and the duration of the survey serve as proxies for effort. We condition on interpretable covariates, including habitat proportions (mixed forest, deciduous forest, shrubs, grass and wetland, agricultural, barren, urban, water body, and coastal), forest stand age, respective volume of pine, spruce, birch, and other deciduous species, vegetation moisture index, and time of year. 
\cite{piirainen2023species} modeled the 120 most common species in two stages, first modeling presences and absences using a probit GLLVM and then modeling the logarithm of positive counts using a Gaussian latent variable model. Their analysis was focused exclusively on predicting species abundances at future times rather than interpreting inferred species distributions and environmental gradients.  

\subsection*{{Scientific Questions:}}
Our primary interest lies in identifying community-level spatial and environmental drivers of avian abundance and using the drivers to infer interpretable community structure. 
\begin{enumerate}
    \item What are the most dominant spatial and environmental drivers that structure Finnish avian communities? The abiotic factors that enable or prevent the establishment or persistence of species are also called ‘environmental filters’, and their identification is one of the most fundamental questions in biodiversity research \citep{kraft2015community}.
    \item How stable are such spatial and environmental drivers over time? Predicting how species respond to the ongoing global change is critically important but highly challenging as current drivers of species distributions may not explain future changes \citep{piirainen2023species}.
    \item What is the structure of species niches over the avian community in terms of how species vary in their level of specialization to particular spatial and environmental drivers? Variation among species in specialization to particular habitats is one of the most fundamental mechanisms behind the co-existence of species in natural communities \citep{buchi2014coexistence} and influences how communities respond to disruptions such as habitat loss and fragmentation \citep{devictor2008distribution}.
    \item To what extent can Finnish avian species be clustered into groups of species with similar environmental and spatial responses? Grouping ecologically similar species aids understanding of ecological complexity \citep{dunstan2011model} and selecting indicator species for ecological monitoring \citep{carignan2002selecting}.
    \item To what extent do the identified spatial and environmental drivers robustly explain the observed avian species community? The reason why such factors may explain only a minority of observed variation is that species distributions are also influenced by a myriad of other ecological and evolutionary assembly processes \citep{weiher2011advances}.
    \item What are the regions with a common avian profile in Finland? Partitioning the environment into areas with similar biological content is useful not only to learn about fundamental ecological questions, but also to help guide resource conservation and utilization \citep{foster2013modelling}.
\end{enumerate}
We endeavor to answer each of these questions using \texttt{barcode}. A complement of six factors (ordination axes), which reflect the dominant drivers of community structure, are identified and related to geography, climate, and habitat. We let the factors vary freely over time and scrutinize resulting patterns to answer Question 2. Species' specializations to the various drivers are addressed by studying each species' preference toward each factor, and we propose clusters based on species-specific latent binary barcodes. We take a model checking approach to answering Question 5: an acceptable model must capture important features of the data without overfitting but can be expected to fit better to some species than others. Lastly, we classify sites based on learned properties of corresponding samples to propose regions of common profile. The identified common regions carry ecological significance and can be interpreted from the lenses of sites and species using inferences from Questions 1–5.

\section{\texttt{barcode} modeling framework}
\label{sec:model}

We adopt the following factorization model for ordination, clustering, and reduced-rank regression. Extending Poisson-gamma matrix factorization, we model factor scores and factor loadings using a combination of continuous and binary latent variables. Sample-specific binary feature vectors indicate the presence of unobserved habitat and sampling conditions, and species-specific binary vectors indicate each species' preference towards the learned habitat factors. The observed covariates and the geographic context are introduced hierarchically to facilitate the interpretation of the learned factors. 

For samples $i=1,2,\dots,n$ and species $j=1,2,\dots, p$, let $y_{ij}\in\{0,1,2,\dots\}$ denote the observed abundance of species $j$ in sample $i$. A sample represents a single line transect survey conducted at one of the $m=555$ survey sites; the site at which the sample $i$ was collected is denoted by $k_i\in\{1,2,\dots,m\}$, and the location of the site $k_i$ by $\mathbf{s}_{k_i}$. Each site was visited between one and eleven times during the study period. Let $x_i=(x_{i1},\dots,x_{iq})^\top$ denote the $q=21$ observed covariates collected for sample $i$. The intercept is included, so $x_{i1}=1$ for all $i=1,\dots,n$.

\subsection{Abundance model}

Observed abundances are conditionally Poisson and independent, and the mean is endowed with a factor-analytic structure,
$$\left(y_{ij}\mid \mu_{ij}\right) \sim \text{Pois}\left(\mu_{ij}\right),\quad 
\mu_{ij}= \theta_i^\top\lambda_j \geq 0,$$
where $\theta_i=\left(\theta_{i1}, \dots,\theta_{iL}\right)^\top$ are latent factor scores describing sample $i$ and $\lambda_{j}=\left( \lambda_{j1},\dots,\lambda_{jL}\right)^\top$ are factor loadings, which capture species-specific preferences for each of the $L$ factors. Univariate gamma priors on both factor scores and factor loadings yields the Bayesian nonnegative matrix factorization or gamma-Poisson factorization model \citep{cemgil2009bayesian, gopalan2015scalable}. Motivated by the importance of simple-to-interpret latent structure and robustness to zeros and large counts, \texttt{barcode} deviates from gamma-Poisson factorization by introducing exact sparsity in factors and loadings,
$$
    \mu_{ij} = (\mathbf{c}_i \circ \phi_i)^\top (\mathbf{s}_j\circ \gamma_j) = \sum_{l=1}^L (c_{il}\phi_{il}) \times (s_{jl}\gamma_{jl}),
$$
where $c_{il}\in \{0,1\}$ and $s_{jl}\in\{0,1\}$. Each factor $\theta_{il}=c_{il}\phi_{il}$ is decomposed into a binary ``switch'' $c_{il}$ and continuous ``strength'' $\phi_{il}$. Loadings $\lambda_{jl}= s_{jl}\gamma_{jl}$ are decomposed similarly with switch $s_{jl}$ and strength $\gamma_{jl}$. We refer to $C\circ \Phi = [c_{il}\phi_{il}]$ as sample factors and $S\circ \Gamma = [s_{jl}\gamma_{jl}]$ as species preferences. The expected abundance for all samples and species is
$$
\mathbb{E} Y = (C\circ \Phi)^\top (S\circ \Gamma)\equiv M.
$$
Together, the sample factors and species preferences determine the mean. Conceptually, the abundance accumulates with each additional factor that is present in the sample and preferred by the species - from a baseline of 0, for each $l$ such that $s_{jl}=1$ \textbf{and} $c_{il}=1$, the expected abundance of species $j$ in sample $i$ increases by $\gamma_{jl}\phi_{il}$, and for each $l$ such that $s_{jl}=0$ \textbf{or} $c_{il}=0$, the expected abundance is unchanged. The first factor is constant; for $l=1$, we fix $c_{il}=1$ and $\phi_{il}=1/n$ for all $i=1,\dots,n$, and we refer to this first factor dimension as the reference factor.

Species-specific loading strengths are assigned a  gamma prior,
$   \gamma_{jl}\sim\text{Ga}\left( a_\gamma, \nu_l \right),$  $\nu_l\sim \text{Ga}\left(a_\nu, b_\nu \right),
$
where $\mathbb{E}_{x\sim \text{Ga}(a,b)} [x] = a/b$. 
To resolve the scale ambiguity between $(\mathbf{c}_l\circ \phi_l)$ and $(\mathbf{s}_l\circ \gamma_l)$, we adopt a prior for $\Phi$ with constrained support. Following \cite{koslovsky2023bayesian}, we define sample intensities through auxiliary variables $\zeta_{il}$ as
\begin{equation}
\begin{aligned}
    \phi_{il} &= \frac{\zeta_{il}}{\sum_{i=1}^n \zeta_{il} c_{il}}, \quad\quad\quad\quad   \zeta_{il}&\sim \text{Ga}\left( \alpha, 1 \right), \label{eq:zetaIntro}
\end{aligned}    
\end{equation}
for $l>1$, which ensures $\mathbf{1}_n^\top (\mathbf{c}_l\circ \phi_l) =1$. This induces the prior
\begin{align}
    \{\phi_{il} : c_{il}=1, i=1,\dots, n\} \sim \text{Dir}\left(\alpha,\dots, \alpha\right), \label{eq:EffDir}
\end{align}
which has effective dimension $\sum_{i}c_{il}\leq n$. 

We find that $\alpha=1$ performs well in a variety of settings. Because the scale of factors is fixed, that is, $\mathbf{1}_n^\top (\mathbf{c}_l\circ \phi_l) =1$, we choose priors for $\nu_l$ that are flexible and computationally convenient as opposed to informative. We set $a_\gamma=0.5$, $a_\nu=0.5$, and $b_\nu=0.5$; these do not depend on $n$, and the marginal expectation and variance of $\gamma$ are unbounded \textit{a priori}. The factorization rank $L$ 
is chosen to achieve good fit to the data while facilitating interpretation.
For the primary results presented below, seven factors are used. Using slightly more or less factors does not alter the dominant trends, as shown in \ref{sec:extras}.  

\subsection{Priors on factor sparsity}
With appropriate priors on $\Pr(c_{il}=1)$ and $\Pr(s_{jl}=1)$, the above model performs unconstrained ordination. This is useful for exploratory analysis when covariates are unavailable. We propose using covariates and spatial temporal context to aid in the interpretation of learned factors and studying the dependence between species abundance and abiotic conditions. Recall that $\mathbf{s}_{k_i}$ denotes the location of the site corresponding to the $i$th sample. The presence of each factor is modeled as a function of covariates and site location:
\begin{align}
    &\text{Pr}\left(c_{il}=1\right) =\begin{cases}
        1 & l=1  \\
        \Phi^{-1}\left( \mathbf{x}_i^\top \beta_l + \xi_{l}\left( \mathbf{s}_{k_i} \right) \right)  & l=2,\dots,L
    \end{cases} \label{model:covariates}\\
    & \xi_l(\mathbf{s}) \sim \mbox{GP}(\mathbf{0}, K) \quad\quad\quad\quad \beta_l \sim \text{N}\left(\mathbf{0}, \sigma^2_0 \text{I}_q\right) \nonumber
\end{align}
where $\xi_l$ is a smooth spatially variable factor-specific intercept shift and $\beta_l=\left(\beta_{l1}, \dots\beta_{lq}\right)^\top $ characterize the effects of covariates on the presence of factor $l$. Being site-specific, the spatial random effect is constant across sampling events. We let $\text{Pr}(s_{jl}=1)=\psi\sim\text{Beta}(10,10)$ \textit{a priori}. We adopt a Gaussian process for $\xi$ using an exponential covariance with length-scale chosen such that the effective range is approximately equal to the 5\% quantile of observed distances. The regression specification can be adapted depending on the context. In simulations studying $\beta$ recovery, we exclude $\xi$.

\subsection{Properties of \texttt{barcode}}\label{sec:properties}

\texttt{barcode} performs unconstrained and concurrent ordination, accommodates extreme sparsity and large counts, clusters samples by habitat profile and species by habitat preference, and can characterize a wide variety of marginal distributions. Here, we provide additional details about these properties as well as other inferential considerations.

\noindent{\textbf{Clustering:}} 
\texttt{barcode} incorporates binary and continuous sample- and species-specific latent variables within a generative probability model for the multivariate abundance counts. These latent variables can be used to define interpretable clusters in a number of ways. As with traditional ordination scores, the continuous loadings order samples and species along $L$ environmental gradients. Through post-processing or post-hoc clustering, the orderings can be translated into clusters. This has been studied in the context of nonnegative matrix factorization, to which \texttt{barcode} is closely related. \cite{kim2008sparse} propose assigning cluster labels based on the index of the largest element of the estimated score vector. In our context, this strategy corresponds to assigning species $j$ to cluster $h_j$, where $h_j=\text{argmax}\ \bm{\gamma}_j$ and sample $i$ to cluster $h_i$, where $h_i=\text{argmax}\ \bm{\phi}_i$. We adopt a variation of this approach to identify regions of common profile. 

Binary barcodes provide an alternative, automatic clustering mechanism. Sample and species barcodes are restricted to lie in $\{0,1\}^{L-1}$ and $\{0,1\}^{L}$ respectively, so unique barcodes can be treated as cluster labels. The possible number of clusters is controlled by $L$, but the number of occupied clusters depends on the level of heterogeneity in the data. Some clusters carry additional interpretation. One-hot clusters (barcodes with only one active factor or preference) identify samples with specialized habitat or climate and species that require a specialized habitat or climate. We use barcodes to cluster species.

\noindent{\textbf{Intercepts:}} 
Two forms of intercept are present in \texttt{barcode}, species intercepts and factor intercepts. Because the reference factor is constant, $c_{i1}\phi_{i1}=1/n$ for all $i$, the preference $s_{j1}\gamma_{j1}$ reflects the expected abundance of species $j$ in the absence of habitat features preferred by the species (i.e., factors indexed by $l>1$). This is a species-specific intercept. Notably, a species' intercept can be exactly zero, signaling its need for specialized habitat. Similarly, a species which loads predominantly on the intercept factor is a generalist with respect to other learned factors. Factor intercepts $\{\beta_{1l}\}_{l>1}$ reflect how common the $l$th factor is among all samples. 

\noindent{\textbf{Interspecific Correlation:}} 
\texttt{barcode} models expected abundance as a function of only latent variables. Therefore, unlike many popular JSDMs based on GLLVMs, there is not an explicit distinction between interspecific marginal correlation due to shared responses to observed environmental covariates and interspecific correlation due to unobserved dynamics (residual correlation). However, diverse patterns of positive and negative interspecific correlation can be well represented by the model. The marginal covariance is
\begin{align}
    \text{Var}(y_{i1},\dots,y_{ip}|S, \Gamma) &= \text{diag}\left( [S\circ\Gamma]\mathrm{E}_i^\top \right) + [S\circ\Gamma] \mathrm{V}_i [S\circ\Gamma]^\top, \nonumber
\end{align}
where $\mathrm{E}_i = \mathbb{E}[ \bm\phi_i\circ\mathbf{c}_i]$ and $\mathrm{V}_i=\text{Var}[ \bm\phi_i\circ\mathbf{c}_i]$. Although species preferences $S\circ\Gamma$ are non-negative, the covariance between factors $V_i$ is unrestricted. Marginal covariance can be negative when factors are negatively correlated. Although factors are uncorrelated \textit{a priori} they can be strongly negatively or positively correlated {\em a posteriori}. See also Supplement \ref{sec:extras}. 

\noindent{\textbf{Rotational Ambiguity:}} 
Factors and loadings from sparse, nonnegative matrix factorizations are potentially subject to non-identifiability due to scale ambiguity and label switching \citep{wang2012nonnegative}. Columnwise normalization of the factor strength matrix $\Phi$ via the Dirichlet prior resolves scale ambiguity. However, the Gibbs sampler may exhibit label switching which, if not corrected, complicates inferences and interpretations. We do not observe this problem here, partly due to the use of factor-specific priors and coordinate-wise updating, and partly to the fact that the learned factors are generally very distinct. Label switching is more likely to arise when the factorization rank is high and factors are positively correlated. When this occurs, samples can be aligned through post-processing \citep{poworoznek2021efficiently}.

\subsection{Posterior Computation}

We employ a straightforward Gibbs sampler for the posterior computation. Here, we briefly describe the data augmentation strategies used, the sampling procedure, and additional considerations. Central to our posterior computation strategy is the fact that the sum of independent Poisson RVs is a Poisson RV. The proposed model can therefore be re-expressed in terms of factor-specific counts $y_{ijl}$, where
\begin{align}
    y_{ijl} \sim \text{Pois}\left( c_{il}\phi_{il}s_{jl}\gamma_{jl} \right),\quad 
    y_{ij} = \sum_{l=1}^L y_{ijl}.  \label{eq:multinomAugment}
\end{align}
Marginally with respect to $\{y_{ijl}\}_{l=1}^{L}$, this preserves $y_{ij} \sim\text{Pois}\left(\mu_{ij}\right)$. Given the factor-specific counts, $\gamma_{jl}$ and $\phi_{il}$ have standard conditional posteriors, and $\left(y_{ij1},\dots,y_{ijL}\mid - \right)$ follows a multinomial distribution. This augmentation has seen widespread use \citep{dunson2005bayesian, cemgil2009bayesian,  gopalan2015scalable}. Coupled with the fact that $\sum_{i} c_{il}\phi_{il}=1$, the additive property of Poisson RVs additionally implies
\begin{align}
    (y_{1jl},\dots,y_{njl}\mid \text{--})\sim \text{Mult}\left( y_{\cdot j l}, \left\{\frac{c_{il}\zeta_{il}}{\sum_ic_{il}\zeta_{il}}\right\}_{i=1}^n\right). \nonumber
\end{align}
Following \cite{koslovsky2023bayesian}, this allows us to introduce 
\begin{align}
    u_l\sim \text{Ga}\bigg( \sum_{i,j} y_{ijl}, \sum_i c_{il}\zeta_{il} \bigg)
    \nonumber 
\end{align}
as a proxy for $\theta_{\cdot l}=\sum_{i}c_{il}\zeta_{il}$, thus affording simple updates for $\zeta_{il}$. The resulting sampler cycles through the following steps:
\begin{enumerate}
\item \textit{Sample binary preferences $s_{jl}$, $j=1:p$ and $l=1:L$ from conditionally independent posteriors, where} 
\begin{align*}
    \text{Pr}(s_{jl}=0\mid \text{--}) \propto& \psi \prod_{i=1}^n \left[\mu^{(0)}_{ij}\right]^{y_{ij}} \exp\left[-\mu^{(0)}_{ij}\right], && \mu^{(0)}_{ij} = \sum_{l'\neq l} c_{il'}\phi_{il'}s_{jl'}\gamma_{jl'}\\
    \text{Pr}(s_{jl}=1\mid \text{--}) \propto& (1-\psi) \prod_{i=1}^n \left[\mu^{(1)}_{ij}\right]^{y_{ij}} \exp\left[-\mu^{(1)}_{ij}\right], && \mu^{(1)}_{ij} = \mu^{(0)}_{ij} + c_{il}\phi_{il}\gamma_{jl}
\end{align*}

\item \textit{Sample binary factors $c_{il}$, $i=1:n$ and $l=1:L$ from conditionally independent posteriors, where} 
\begin{align*}
    \text{Pr}(c_{il}=0\mid \text{--}) \propto& \Phi^{-1}\left(\mathbf{x}_i^\top \beta_l + \xi_{lk_i}\right) \prod_{j=1}^p \left[\mu^{(0)}_{ij}\right]^{y_{ij}} \exp\left[-\mu^{(0)}_{ij}\right], \\
    & \mu^{(0)}_{ij} = \sum_{l'\neq l} c_{il'}\phi_{il'}s_{jl'}\gamma_{jl'}\\
    \text{Pr}(c_{il}=1\mid \text{--}) \propto& \left[1-\Phi^{-1}\left(\mathbf{x}_i^\top \beta_l + \xi_{lk_i}\right)\right]\psi \prod_{i=1}^n \left[\mu^{(1)}_{ij}\right]^{y_{ij}} \exp\left[-\mu^{(1)}_{ij}\right], \\
    & \mu^{(1)}_{ij} = \mu^{(0)}_{ij} + \phi_{il}s_{jl}\gamma_{jl}
\end{align*}

\item \textit{Sample factor-specific counts $\left(y_{ij1}, \dots, y_{ijL}\right)$, $i=1:n$ and $j=1:p$}
\begin{align*}
    \left(y_{ij1}, \dots, y_{ijL}\mid \text{--}\right) &\sim \text{Mult}\left( y_{ij}, \pi_{ij} \right), \quad\quad\quad\quad \pi_{ijl} = \frac{c_{il}\phi_{il}s_{jl}\gamma_{jl}}{\sum_l c_{i1}\phi_{i1}s_{j1}\gamma_{j1}}
\end{align*}
\item \textit{Sample preference intensities $\gamma_{jl}$, $j=1:p$ and $l=1:L$}:
$$
    \left(\gamma_{jl}\mid \text{--} \right) \sim \text{Ga}
    \left( a_\gamma +  y_{\cdot jl}, \nu_l + 1 \right) 
$$
\item \textit{Sample auxiliary normalizing constants $u_l$, $l=1:L$}:
\begin{align*}
    (u_l\mid \text{--})\sim \text{Ga}\bigg( y_{\cdot \cdot l}, \sum_i c_{il}\zeta_{il} \bigg)
\end{align*}
\item \textit{Sample unnormalized factor intensities $\zeta_{il}$, $i=1:n$ and $l=1:L$}:
\begin{align*}
    (\zeta_{il}\mid \text{--}) &\sim \text{Ga}\left( \alpha + y_{i\cdot l}, 1 + u_l \right)
\end{align*}
\item \textit{Sample latent regression coefficients and random effects $\beta$, $\xi$ following \cite{albert1993bayesian}.}
\item \textit{Sample hyperparameters $\psi$ and $\nu_l$, $l=1:L$ from conditionally independent posteriors. }
\end{enumerate}
The sampler is generally efficient. A key source of this efficiency is that sampling factor-specific latent counts \eqref{eq:multinomAugment} scales with the number of non-zero elements of $Y$. We employ three simple strategies to further improve efficiency: 1) a pseudo-prior for $\gamma_{jl}$, 2) jointly updating blocks of binary latent variables to encourage exploration of the discrete space, and 3) a warm start to limit the influence of multimodality inherent to Poisson factorization models. These strategies are detailed further in Supplement \ref{sec:CompDetails}.

\section{Factors Driving Finnish Bird Abundance Data}\label{sec:results}

We fit \texttt{barcode} with seven factors, including the reference. We find that seven factors reliably capture dominant trends without resulting in factor splitting and highly correlated, difficult-to-interpret factors. However, results using slightly more and fewer factors, which are included as supplementary material, support the general results we now present. The learned factors reflect the dominant axes of variation in Finnish avian communities. We first study the extent to which these factors can be explained by known habitat and climatic gradients, which leads to informative and intuitive interpretations of each. The factors are used to infer preferences, specializations, and clusters of avian species. Together, inferences on factors and inferences on corresponding species preferences highlight regions of common avian community profile. 

\subsubsection*{Dominant Spatial and Environmental Drivers}
\begin{figure}[h]
\includegraphics[width=1.055\textwidth, center]{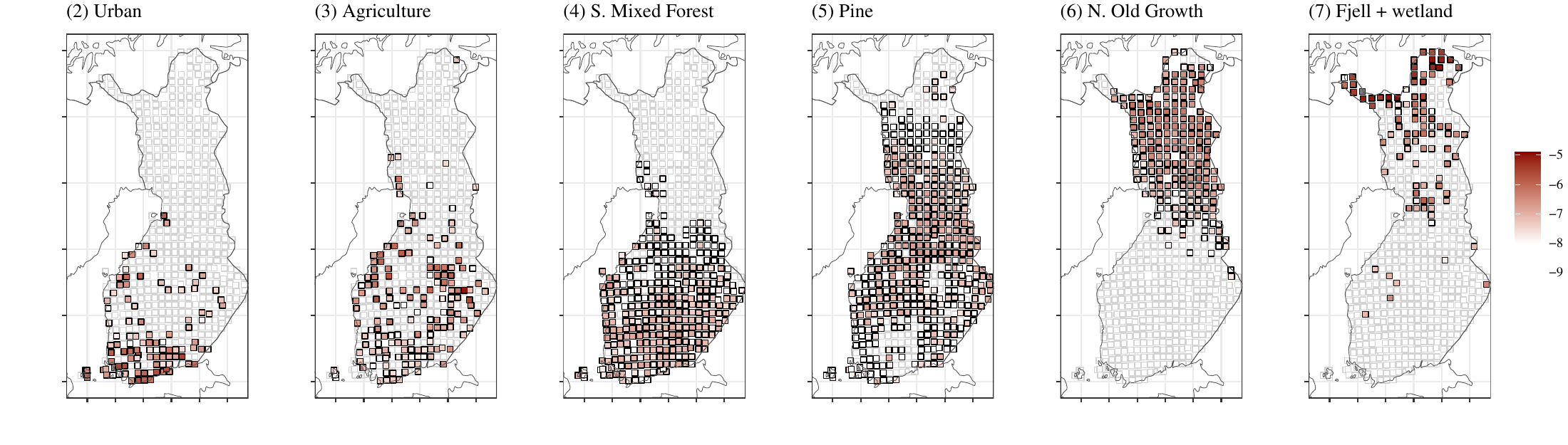}
\caption{\label{fig:maps7} Factor presence and strength overlaid on the map of Finland. Sampling sites are shown as tiles. Black borders indicate the presence of each factor, and coloration indicates log-transformed factor strength. For sites that are sampled multiple times, border and color values represent averages over samples within each site.}
\end{figure}

To aid in interpretation, factors were reordered post hoc. We refer to a factor as ``present'' when $c_{il}=1$ with the understanding that factor dimensions are latent statistical constructs to which we assign ecological meaning. Posterior medians of $C$ and $S$ indicate that each factor is present in (100\%,  28\%,  43\%,  63\%,  68\%,  34\%, and  19\%) of samples, respectively, and (41\%, 67\%, 57\%, 64\%, 62\%, 52\%, 29\%) percent of species prefer each corresponding factor. Figure \ref{fig:maps7} displays the map of Finland with colored tiles indicating the presence of the factors at each site, averaged over the samples taken at each site and colored by the estimated factor strength. 

\subsubsection*{Habitat and Climatic Variation of Factors}

Using this decomposition, we identify three factors that smoothly segment the region by geography (especially climatic gradient) and forest type, and three factors that indicate specialized habitat types. Factors 4 and 6 are negatively correlated and define southern Finland (with benign climatic conditions) and northern Finland (with harsh climatic conditions), respectively. Only 4\% of the samples scored positively in both 4 and 6, although 39 species (30\%) prefer both. Factor 5 interpolates factors 4 and 6: it is strongest at mid-latitudes and weakest at extreme latitudes, and all but three of the species that load on 4 and 6 also load on 5. In contrast, factors 2, 3 and 7 do not exhibit clear and smooth geospatial gradients but rather clusters or isolated sites. Factor 2 is isolated to south latitudes and is present at all sites along the South Coast. The South Coast hosts a large proportion of the country's human population. Other sites where factor 2 is present also align with human demographic patterns in Finland; population centers removed from the South Coast including Oulu, Joensuu, and Vaasa are co-located with factor 2 sites. Factor 3 similarly exhibits geospatial heterogeneity but correspondence to known geographic or demographic patterns is not immediately obvious. Finally, factor 7 is present in all northernmost sites, although this does not fully explain its distribution. 

\begin{figure}
\includegraphics[width=0.8\textwidth, center]{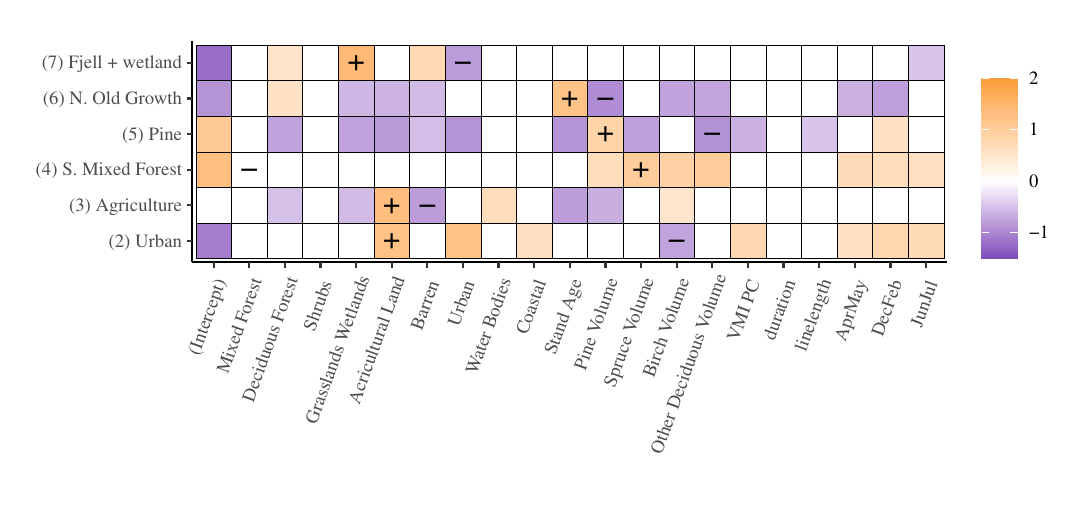}
\caption{\label{fig:cov7} Effects of covariates on binary sample factors $C$. Colored tiles indicate effects that are strictly positive or negative with high posterior probability ($>0.95$). Plus signs (resp. minus signs) identify the strongest positive (resp. negative) effect for each factor. Note that for factor 4, the leading negative effect is not well-separated from zero. }
\end{figure}

Despite compelling visual patterns, factor interpretations should be made carefully, as patterns in latent factor occurrence are due to both observed and unobserved environmental gradients. We estimated the effects of 20 standardized covariates on factor occurrence to refine our interpretations. The covariates include nine measured habitat types, six forest inventory metrics, and five variables that describe the sampling conditions. Figure \ref{fig:cov7} summarizes the effects. Estimated covariate effects on each factor should be interpreted as a summary of the typical sample or site where the factor is present and should be used in conjunction with other evidence to make comprehensive interpretations. Factor 2 occurs most often in conjunction with agricultural, urban, and coastal environments and least often with high-density birch forests. Factor 3 is difficult to interpret based only on its geospatial distribution; the leading covariate effects suggest that factor 3 specifically indicates agricultural open spaces. In addition to representing particular latitude bands, factor 4 is associated with diverse high-volume forests, while factor 5 is associated specifically with pine-dominated forests. Factor 6, which is only found in northern Finland, is associated with old-growth forests, specifically excluding pine. The association between factor 7 and grasslands, wetlands, and barren areas is consistent with its appearance in the northern reaches of Finnish Lapland and suggests that the sites hosting factor 7 in the south are likely uncultivated open spaces and wetlands. Additionally, the negative effect of urban habitat is consistent with the low population density at high latitudes and the co-location of national parks and wilderness areas with northern sites. 

Together, geospatial distributions and covariate effects suggest the following interpretations of spatio-environmental drivers (short names used in figures are given in parentheses):
1. (Reference), 
2. Human-built environment (Urban), 
3. Agricultural areas (Agriculture), 
4. Mixed forest in Southern Finland (S. Mixed Forest), 
5. Managed pine forest (Pine),
6. Old-growth forest in Northern Finland (N. Old Growth),
and 7. Arctic fjells and wetlands (Fjell + Wetland).

\subsubsection*{Temporal Variation of Factors}

\begin{figure}
\includegraphics[width=0.7\textwidth, center]{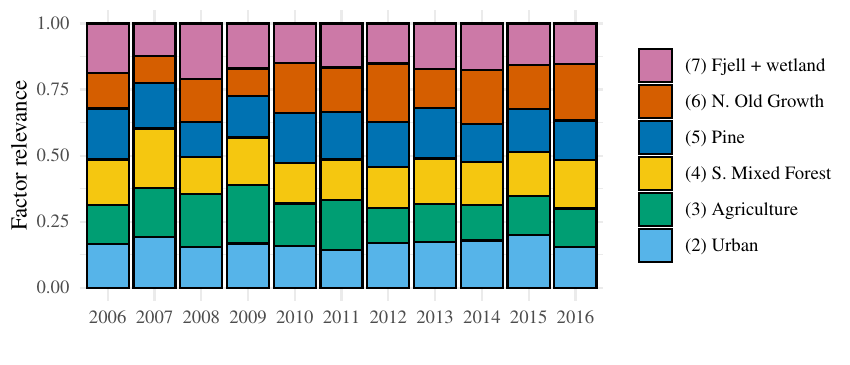}
\caption{\label{fig:time7} Relative cumulative factor strengths in each year of the study.}
\end{figure}

When a factor is present (or absent) from a sample, it is usually the case that the same factor is present (or absent) from other samples taken at that site. The fractions of sites for which all corresponding sample switches are 0 or 1 are (100\%,  91\%, 79\%,  90\%,  85\%, 90\%,  90\%), respectively. Factor 3, agricultural areas across Finland, shows the lowest site-factor fidelity. A possible explanation is that agricultural habitats are intrinsically more unstable than some others given that fields may be regularly replanted, rotated, or left fallow, and each option leads to potentially significantly different resource availability. Figure \ref{fig:time7} displays the relative importance of each factor per year. The small fluctuations between years may be attributed partly to annual variation in climatic conditions and partly to the differences in sampling effort between years. Although some sites were sampled all or most years, most sites were sampled fewer than five times, and the total number of samples in a given year varied between 205 and 300. If sites hosting some factors are systematically undersampled in some years, it is natural to expect variation in the year-over-year distribution of factors.

\subsubsection*{Species Specialization to Learned Drivers}

We next investigate species' relationships to these factors and one another. Some species load on several factors, whereas others specialize by loading on one factor only. Specialists are interesting because their identities can further confirm or contradict proposed factor interpretations, and, by nature of specialization, may be more susceptible to habitat disturbance or loss. We find that only factors 1, 2, 6, and 7 host specialists. Urban and coastal specialists (factor 2) are the common pigeon (\textit{Columba livia}), Eurasian oystercatcher (\textit{Haematopus ostralegus}), mute swan (\textit{Cygnus olor}), and European goldfinch (\textit{Carduelis carduelis}). The little bunting (\textit{Emberiza pusilla}) and Lapland longspur (\textit{Calcarius lapponicus}) specialize in northern/old-growth (6) and Arctic fjells/wetland (7), respectively. The little bunting typically breeds in open coniferous peatlands in the north and agricultural environments rather than in old-growth forests. This highlights the importance of using multiple lines of evidence to interpret factors--the northern climate indicated by factor 6 is more relevant than old-growth forests in the case of little bunting's distribution. Factors 3, 4, and 5 do not have single-factor specialists. Willow ptarmigan (\textit{Lagopus lagopus}), broad-billed sandpiper (\textit{Calidris falcinellus}), and bluethroat (\textit{Luscinia svecica}) prefer only factors 6 and 7; common starling (\textit{Sturnus vulgaris}), great crested grebe (\textit{Podiceps cristatus}), and house sparrow (\textit{Passer domesticus}) prefer only 2 and 3; five species prefer only 2 and 4. 

\begin{figure}
\begin{center}
\includegraphics[width=1\textwidth]{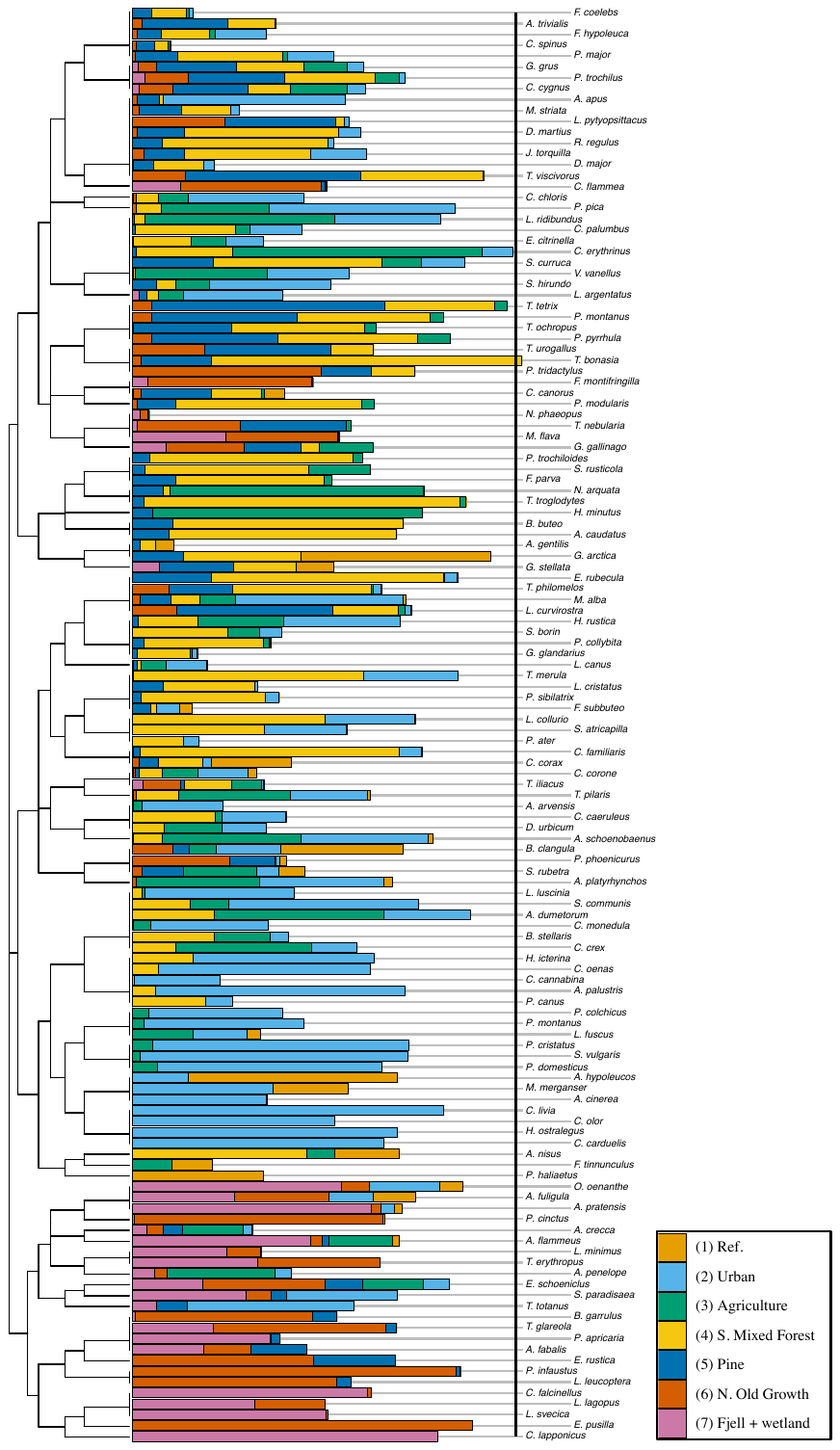}
\end{center}
\caption{\label{fig:vertClustersExport} Species profiles clustered by preference and segmented by preference strength. For each species, the width of each bar indicates the posterior marginal variance relative to the empirical marginal variance. The vertical black line denotes equality between posterior and empirical marginal variance. In some cases (e.g., \textit{Carpodacus erythrinus} and \textit{Tetrastes bonasia}), the learned factors explain much of the observed variation, whereas for many others, the factors are less explanatory.}
\end{figure}

The reference factor (1) is constant across sites and years, so loading on the reference identifies a species as a generalist, or as occurring in patterns that are not well explained by other factors. Species which load strongly on the reference include several ducks and waterbirds such as the common merganser (\textit{Mergus merganser}), common goldeneye (\textit{Bucephala clangula}), black-throated loon (\textit{Gavia arctica}), and common sandpiper (\textit{Actitis hypoleucos}). Only ospreys (\textit{Pandion haliaetus}) specialize to the reference, which further supports its interpretation as identifying generalist species—ospreys are one of the very few bird species known to have a global distribution \citep{monti2015being}. We note that all these species require the presence of aquatic habitats, but given the abundance of lakes and ponds in Finland, such habitats are likely to be present along the majority of the transect lines.

\subsubsection*{Clustering Species by Preferences}

Manual factor-wise partitioning of species is useful but quickly becomes intractable for even a small number of factors, as most species load on several factors. A defining feature of \texttt{barcode} is the introduction of binary latent variables $\bm{s}_j$, which provide a simple index by which to cluster species that can be mapped back to interpretable habitat factors. The matrix $S$ implicitly groups species in up to $2^7-1=127$ clusters of which only 56 are occupied, and 25 of which are occupied by only one species. Figure \ref{fig:vertClustersExport} adds nuance to these clusters in three ways: 1) reordering species by hierarchical clustering of rows of $S$, 2) displaying the fraction of variance explained, and 3) indicating the relative importance of each factor to each species. The hierarchical clustering distills the large number of clusters into more interpretable groups; although many clusters contain only one species, they can be grouped further. The fraction of variance explained, which is calculated as the posterior marginal variance divided by the empirical marginal variance, places interpretations within the context of model fit. Finally, the relative factor importance uses $\Gamma$ to tease apart preferences and preference strengths.  

We find that mixed and managed southern pine forests are simultaneously preferred by many species but explain their distributions to varying degrees—mistle thrush (\textit{Turdus viscivorus}) and great spotted woodpecker (\textit{Dendrocopos major}) share identical barcodes, but variation in the distribution of mistle thrush is much better explained. Figure \ref{fig:vertClustersExport} also highlights the importance of factor pairs: 2 and 3 (urban and agricultural), 6 and 7 (northern old growth and wetland/fjell), and 4 and 5 (mixed and pine forests), which agrees with intuition. A notable exception to the urban-agricultural coupling are predatory species: the three fully predatory raptor species that prefer agricultural environments avoid urban areas during the breeding season represented by our data (Eurasian sparrowhawk, \textit{Accipiter nisus}; common kestrel, \textit{Falco tinnunculus}; short-eared owl, \textit{Asio flammeus}), as do all specialized invertebrate predators (common snipe, \textit{Gallinago gallinago}; Eurasian curlew, \textit{Numenius arquata}; Eurasian woodcock, \textit{Scolopax rusticola}). The species least explained by our modelling approach are Eurasian siskin (\textit{Carduelis spinus}), Eurasian whimbrel (\textit{Numenius phaeopus}), and Eurasian goshawk (\textit{Accipiter gentilis}).

\subsubsection*{Explaining and Predicting Observed Community Distribution}

We study species-by-species summaries to assess fit and out-of-sample prediction to assess generalizability. For all species, the ratio between posterior predictive and empirical marginal expectation is very close to one (\ref{fig:margMean}), indicating that the learned factors effectively characterize each species' average abundance. Figure \ref{fig:vertClustersExport} compares marginal variances and highlights that while the variation in the distribution of some species is well explained, others are more variable than expected by the model. 

With sufficiently many factors, factor models such as \texttt{barcode} and GLLVMs can fit the data arbitrarily well, but overfitting is a concern.
Predicting unseen data is useful for identifying both lack of fit and overfitting. Therefore we use three-fold cross validation to evaluate \texttt{barcode} and compare to Poisson and NB GLLVMs. Details, including the number of species- and sample-specific parameters required by each model, are provided in the supplement. The out-of-sample RMSE averaged over folds using \texttt{barcode} with four, seven, and ten factors was 3.84, 3.76, and 3.73, respectively. Using Poisson and NB GLLVM, it was 47.95 and 4392.19. These large values reflect the fact that the log-linear models sporadically generate predictions far outside reasonable limits when simultaneously confronted with sparsity and large counts. Identifying outlying predictions is not straightforward because, although some large predictions are obviously errant (e.g., 10000 times larger than the largest observed count), some are plausible (e.g., 2 times larger than the largest observed count). When predictions made by Poisson and NB are filtered to exclude those at least one order of magnitude greater than all data, the average RMSEs are 4.39 and 3.66. These results indicate that \texttt{barcode} fits the joint distribution of data comparably to standard GLLVMs without predicting extreme values. This also suggests that a small improvement in fit and out-of-sample prediction can be achieved by increasing the number of factors from seven to ten. However, this small improvement is not worth the more complex interpretation.

\subsubsection*{Regions of Common Profile}

\begin{figure}
\begin{center}
\includegraphics[width=0.45\textwidth]{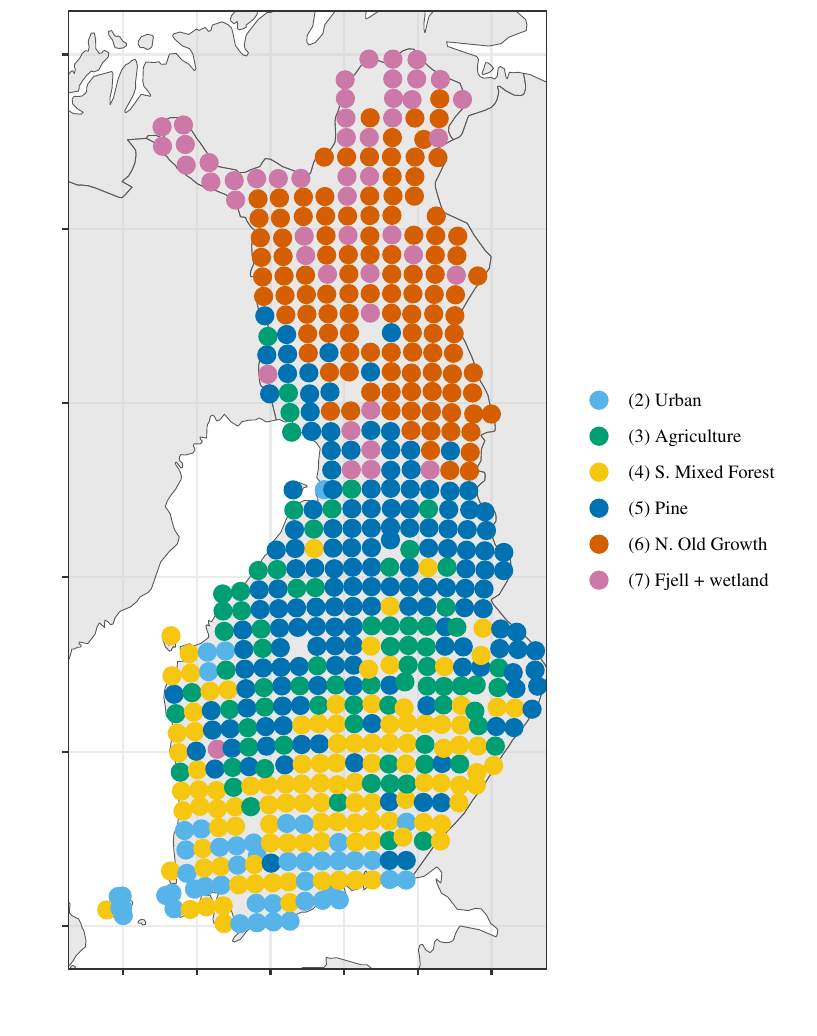}
\end{center}
\caption{\label{fig:RCP} Regions of Finland sharing common avian community profiles. Points are colored by the dominant factor in each site, over all samples from the site. }
\end{figure}

Having characterized dominant environmental factors and the ways in which they can be used to infer community structure, we propose using the factors to define regions having a common avian profile. We classify each site by the factor that is dominant on average over samples from the site. This leads to an interpretable decomposition of Finland into coarse habitat and climatic regions, which are directly mapped to species preferences (Figure \ref{fig:RCP}). Latitudinal trends are immediately discernible, as well as regions of isolated urban, mixed forest and wetland habitat, which suggests the design of broad regional management plans and targeted local plans. 

This classification, based on the most dominant factor at each site, is but one way to define regions of common profile. A more precise classification can be obtained by using the barcode-based clustering procedure, which would lead to up to $2^k - 1$ distinct classes similar to the species clusters presented previously.

\section{Discussion}

We introduced an approach to unconstrained and concurrent ordination that utilizes an additive Poisson model and discrete latent variables to learn interpretable factors driving variation in observed abundance data. The model structure enables community-level inference through simultaneous ordination, clustering, and latent regression. Analysis of Finnish avian community data identified six dominant environmental drivers of community structure including three climatic regions dominated by distinct forest types and three spatially heterogeneous drivers corresponding to specialized habitats. The drivers are generally stable year-over-year. We proposed species clusters based on their respective preferences towards each factors, and identified northern old growth and arctic fjell/wetland environments as hosting more specialized species. Finally, we united these inferences to identify regions of common profile across Finland, which include both geographically contiguous and isolated regions. More generally, this analysis expands the suite of models useful for ordination and clustering using abundance data. 

Although the development of this approach is motivated in part by deficiencies of existing multivariate count models, extending to other kinds of ecological data would be both straightforward and useful. Binary presence-absence data $z$ can be modeled as a censored version of a latent count $y$, where $z=\mathbbm{1}(y > 0)$ \citep{dunson2005bayesian, pmlr-v38-zhou15a}. The \texttt{barcode} model can be applied directly to the latent counts. Although interpretations would change slightly, the same qualitative inferences could be drawn. Similarly, given the close correspondence between multinomial and Poisson models for categorical data \citep{forster2010bayesian}, \texttt{barcode} is a promising foundation for compositional data modeling.

While promising, \texttt{barcode} is not without limitations. Our analysis fails to directly account for detection errors. A volunteer or expert observer can be expected to misidentify or fail to detect individuals while sampling. The rates of misidentification and nondetection vary by species and with survey setting (e.g., effort, time of day, weather conditions), which can affect inferences \citep{steenweg2019species, lahoz2014imperfect}. While we remain confident in the community-level trends identified by our analysis, better accounting for nondetection using an occupancy modeling framework \citep{royle2005modelling} is advisable and may lead to better model fit for poorly-explained species (Figure \ref{fig:vertClustersExport}). An alternative strategy is to integrate ideas from occupancy modeling and matrix completion. Presently, \texttt{barcode} treats observed zeros as ``true'' zeros--no individuals of the species are present--which is problematic in the presence of systematic detection errors. In collaborative filtering or recommender system contexts, where nonnegative Poisson matrix factorization also enjoys popularity, this property manifests when the absence of a user rating or interaction is conflated with a zero rating or interaction score \citep{gopalan2015scalable}. \cite{basbug2016hierarchical} confront this issue by separating the model for sparsity from the model for ratings or abundance, while preserving the appealing computational and inferential properties of the additive Poisson model. Adapting this approach to the occupancy modeling setting and incorporating survey effort directly is an interesting future direction. 

In \texttt{barcode}, sample factors freely vary over time. For data that exhibit more temporality, a more structured model for temporal variation may be merited. In discrete time, a tensor factorization could be adopted, where sites, species, and time points each receive factor scores \citep{yoo2009probabilistic, schein2014inferring}. 

Although alternative count models are commonly used for community abundance data, the conditional Poisson specification is surprisingly flexible and allows for efficient model fitting and inference. A popular way to induce a negative binomial or other overdispersed Poisson distribution is as the marginal of a conditional Poisson model. By conditioning the outcome distribution on latent variables and modeling factor and factor loading distributions hierarchically, we induce an intricate Poisson mixture structure.  Nonetheless, it would be natural to consider different distributional families for the response and latent variables. Furthermore, although negative correlations can be represented by the method, they arise as a consequence of the chosen mean model. Thus, the modeler has relatively little control and must trust in the flexibility of the model to accurately capture the joint distribution. One consequence of this, which is shared by standard Poisson and NB GLLVMs, is an unintentional coupling of interspecific correlation and marginal overdispersion--highly correlated species must be marginally overdispersed. More flexible approaches to modeling dependence among species counts are needed. 

\newpage 

\bibliography{refs}

\begin{thebibliography}{67}
\providecommand{\natexlab}[1]{#1}
\providecommand{\url}[1]{\texttt{#1}}
\expandafter\ifx\csname urlstyle\endcsname\relax
  \providecommand{\doi}[1]{doi: #1}\else
  \providecommand{\doi}{doi: \begingroup \urlstyle{rm}\Url}\fi

\bibitem[Ahola et~al.(2007)Ahola, Laaksonen, Eeva, and Lehikoinen]{ahola2007climate}
M.~P. Ahola, T.~Laaksonen, T.~Eeva, and E.~Lehikoinen.
\newblock Climate change can alter competitive relationships between resident and migratory birds.
\newblock \emph{Journal of Animal Ecology}, 76\penalty0 (6):\penalty0 1045--1052, 2007.

\bibitem[Albert and Chib(1993)]{albert1993bayesian}
J.~H. Albert and S.~Chib.
\newblock Bayesian analysis of binary and polychotomous response data.
\newblock \emph{Journal of the American Statistical Association}, 88\penalty0 (422):\penalty0 669--679, 1993.

\bibitem[Basbug and Engelhardt(2016)]{basbug2016hierarchical}
M.~Basbug and B.~Engelhardt.
\newblock Hierarchical compound {P}oisson factorization.
\newblock In \emph{International Conference on Machine Learning}, pages 1795--1803. PMLR, 2016.

\bibitem[B{\"u}chi and Vuilleumier(2014)]{buchi2014coexistence}
L.~B{\"u}chi and S.~Vuilleumier.
\newblock Coexistence of specialist and generalist species is shaped by dispersal and environmental factors.
\newblock \emph{The American Naturalist}, 183\penalty0 (5):\penalty0 612--624, 2014.

\bibitem[Carignan and Villard(2002)]{carignan2002selecting}
V.~Carignan and M.-A. Villard.
\newblock Selecting indicator species to monitor ecological integrity: a review.
\newblock \emph{Environmental Monitoring and Assessment}, 78\penalty0 (1):\penalty0 45--61, 2002.

\bibitem[Carvalho et~al.(2008)Carvalho, Chang, Lucas, Nevins, Wang, and West]{carvalho2008high}
C.~M. Carvalho, J.~Chang, J.~E. Lucas, J.~R. Nevins, Q.~Wang, and M.~West.
\newblock High-dimensional sparse factor modeling: applications in gene expression genomics.
\newblock \emph{Journal of the American Statistical Association}, 103\penalty0 (484):\penalty0 1438--1456, 2008.

\bibitem[Cemgil(2009)]{cemgil2009bayesian}
A.~T. Cemgil.
\newblock Bayesian inference for nonnegative matrix factorisation models.
\newblock \emph{Computational Intelligence and Neuroscience}, 2009\penalty0 (1):\penalty0 785152, 2009.

\bibitem[Chase and Leibold(2009)]{chase2009ecological}
J.~M. Chase and M.~A. Leibold.
\newblock \emph{{Ecological Niches: Linking Classical and Contemporary Approaches}}.
\newblock University of Chicago Press, 2009.

\bibitem[Cortes(2018)]{cortes2018fast}
D.~Cortes.
\newblock Fast non-{B}ayesian {P}oisson factorization for implicit-feedback recommendations.
\newblock \emph{arXiv preprint arXiv:1811.01908}, 2018.

\bibitem[Davies and Tso(1982)]{davies1982procedures}
P.~Davies and M.~K. Tso.
\newblock Procedures for reduced-rank regression.
\newblock \emph{Journal of the Royal Statistical Society Series C: Applied Statistics}, 31\penalty0 (3):\penalty0 244--255, 1982.

\bibitem[Dellaportas et~al.(2002)Dellaportas, Forster, and Ntzoufras]{dellaportas2002bayesian}
P.~Dellaportas, J.~J. Forster, and I.~Ntzoufras.
\newblock On {B}ayesian model and variable selection using {MCMC}.
\newblock \emph{Statistics and Computing}, 12\penalty0 (1):\penalty0 27--36, 2002.

\bibitem[Devictor et~al.(2008)Devictor, Julliard, and Jiguet]{devictor2008distribution}
V.~Devictor, R.~Julliard, and F.~Jiguet.
\newblock Distribution of specialist and generalist species along spatial gradients of habitat disturbance and fragmentation.
\newblock \emph{Oikos}, 117\penalty0 (4):\penalty0 507--514, 2008.

\bibitem[Dunson and Herring(2005)]{dunson2005bayesian}
D.~B. Dunson and A.~H. Herring.
\newblock Bayesian latent variable models for mixed discrete outcomes.
\newblock \emph{Biostatistics}, 6\penalty0 (1):\penalty0 11--25, 2005.

\bibitem[Dunstan et~al.(2011)Dunstan, Foster, and Darnell]{dunstan2011model}
P.~K. Dunstan, S.~D. Foster, and R.~Darnell.
\newblock Model based grouping of species across environmental gradients.
\newblock \emph{Ecological Modelling}, 222\penalty0 (4):\penalty0 955--963, 2011.

\bibitem[Engelhardt et~al.(2020)Engelhardt, Neuschulz, and Hof]{engelhardt2020ignoring}
E.~K. Engelhardt, E.~L. Neuschulz, and C.~Hof.
\newblock Ignoring biotic interactions overestimates climate change effects: The potential response of the spotted nutcracker to changes in climate and resource plants.
\newblock \emph{Journal of Biogeography}, 47\penalty0 (1):\penalty0 143--154, 2020.

\bibitem[Forster(2010)]{forster2010bayesian}
J.~J. Forster.
\newblock Bayesian inference for {Poisson} and multinomial log-linear models.
\newblock \emph{Statistical Methodology}, 7\penalty0 (3):\penalty0 210--224, 2010.

\bibitem[Foster et~al.(2013)Foster, Givens, Dornan, Dunstan, and Darnell]{foster2013modelling}
S.~Foster, G.~Givens, G.~Dornan, P.~Dunstan, and R.~Darnell.
\newblock Modelling biological regions from multi-species and environmental data.
\newblock \emph{Environmetrics}, 24\penalty0 (7):\penalty0 489--499, 2013.

\bibitem[Fr{\"u}hwirth-Schnatter(2023)]{fruhwirth2023generalized}
S.~Fr{\"u}hwirth-Schnatter.
\newblock Generalized cumulative shrinkage process priors with applications to sparse {Bayesian} factor analysis.
\newblock \emph{Philosophical Transactions of the Royal Society A}, 381\penalty0 (2247):\penalty0 20220148, 2023.

\bibitem[Gaujoux and Seoighe(2010)]{gaujoux2010flexible}
R.~Gaujoux and C.~Seoighe.
\newblock A flexible {R} package for nonnegative matrix factorization.
\newblock \emph{BMC Bioinformatics}, 11:\penalty0 1--9, 2010.

\bibitem[Gelman and Rubin(1992)]{gelman1992inference}
A.~Gelman and D.~B. Rubin.
\newblock Inference from iterative simulation using multiple sequences.
\newblock \emph{Statistical Science}, 7\penalty0 (4):\penalty0 457--472, 1992.

\bibitem[Gopalan et~al.(2015)Gopalan, Hofman, and Blei]{gopalan2015scalable}
P.~Gopalan, J.~M. Hofman, and D.~M. Blei.
\newblock {Scalable Recommendation with Hierarchical Poisson Factorization}.
\newblock In \emph{UAI}, pages 326--335, 2015.

\bibitem[Gower(1966)]{gower1966some}
J.~C. Gower.
\newblock Some distance properties of latent root and vector methods used in multivariate analysis.
\newblock \emph{Biometrika}, 53\penalty0 (3-4):\penalty0 325--338, 1966.

\bibitem[Gu and Dunson(2023)]{gu2023bayesian}
Y.~Gu and D.~B. Dunson.
\newblock Bayesian pyramids: {Identifiable} multilayer discrete latent structure models for discrete data.
\newblock \emph{Journal of the Royal Statistical Society Series B: Statistical Methodology}, 85\penalty0 (2):\penalty0 399--426, 2023.

\bibitem[Hoegh and Roberts(2020)]{hoegh2020evaluating}
A.~Hoegh and D.~W. Roberts.
\newblock Evaluating and presenting uncertainty in model-based unconstrained ordination.
\newblock \emph{Ecology and Evolution}, 10\penalty0 (1):\penalty0 59--69, 2020.

\bibitem[Hui(2017)]{hui2017model}
F.~K. Hui.
\newblock Model-based simultaneous clustering and ordination of multivariate abundance data in ecology.
\newblock \emph{Computational Statistics \& Data Analysis}, 105:\penalty0 1--10, 2017.

\bibitem[Hui et~al.(2015)Hui, Taskinen, Pledger, Foster, and Warton]{hui2015model}
F.~K. Hui, S.~Taskinen, S.~Pledger, S.~D. Foster, and D.~I. Warton.
\newblock Model-based approaches to unconstrained ordination.
\newblock \emph{Methods in Ecology and Evolution}, 6\penalty0 (4):\penalty0 399--411, 2015.

\bibitem[Kim and Park(2008)]{kim2008sparse}
J.~Kim and H.~Park.
\newblock Sparse nonnegative matrix factorization for clustering.
\newblock Technical report, Georgia Institute of Technology, 2008.

\bibitem[Koslovsky(2023)]{koslovsky2023bayesian}
M.~D. Koslovsky.
\newblock A {B}ayesian zero-inflated {D}irichlet-multinomial regression model for multivariate compositional count data.
\newblock \emph{Biometrics}, 79\penalty0 (4):\penalty0 3239--3251, 2023.

\bibitem[Kraft et~al.(2015)Kraft, Adler, Godoy, James, Fuller, and Levine]{kraft2015community}
N.~J. Kraft, P.~B. Adler, O.~Godoy, E.~C. James, S.~Fuller, and J.~M. Levine.
\newblock Community assembly, coexistence and the environmental filtering metaphor.
\newblock \emph{Functional Ecology}, 29\penalty0 (5):\penalty0 592--599, 2015.

\bibitem[Kruskal(1978)]{kruskal1978multidimensional}
J.~B. Kruskal.
\newblock Multidimensional scaling.
\newblock \emph{Murry Hill}, 1978.

\bibitem[Lahoz-Monfort et~al.(2014)Lahoz-Monfort, Guillera-Arroita, and Wintle]{lahoz2014imperfect}
J.~J. Lahoz-Monfort, G.~Guillera-Arroita, and B.~A. Wintle.
\newblock Imperfect detection impacts the performance of species distribution models.
\newblock \emph{Global Ecology and Biogeography}, 23\penalty0 (4):\penalty0 504--515, 2014.

\bibitem[Lal et~al.(2021)Lal, Liu, Tibshirani, Sidow, and Ramazzotti]{lal2021novo}
A.~Lal, K.~Liu, R.~Tibshirani, A.~Sidow, and D.~Ramazzotti.
\newblock De novo mutational signature discovery in tumor genomes using {SparseSignatures}.
\newblock \emph{PLoS Computational Biology}, 17\penalty0 (6):\penalty0 e1009119, 2021.

\bibitem[Langville et~al.(2006)Langville, Meyer, Albright, Cox, and Duling]{langville2006initializations}
A.~N. Langville, C.~D. Meyer, R.~Albright, J.~Cox, and D.~Duling.
\newblock Initializations for the nonnegative matrix factorization.
\newblock In \emph{Proceedings of the twelfth ACM SIGKDD international conference on knowledge discovery and data mining}, pages 23--26. Citeseer, 2006.

\bibitem[Lee and Gu(2024)]{lee2024new}
S.~Lee and Y.~Gu.
\newblock New paradigm of identifiable general-response cognitive diagnostic models: beyond categorical data.
\newblock \emph{Psychometrika}, 89:\penalty0 1--33, 2024.

\bibitem[Legendre and Legendre(2012)]{legendre2012numerical}
P.~Legendre and L.~Legendre.
\newblock \emph{{Numerical Ecology}}.
\newblock Elsevier, 2012.

\bibitem[Legramanti et~al.(2020)Legramanti, Durante, and Dunson]{legramanti2020bayesian}
S.~Legramanti, D.~Durante, and D.~B. Dunson.
\newblock Bayesian cumulative shrinkage for infinite factorizations.
\newblock \emph{Biometrika}, 107\penalty0 (3):\penalty0 745--752, 2020.

\bibitem[Lehikoinen et~al.(2014)Lehikoinen, Green, Husby, K{\aa}l{\aa}s, and Lindstr{\"o}m]{lehikoinen2014common}
A.~Lehikoinen, M.~Green, M.~Husby, J.~A. K{\aa}l{\aa}s, and {\AA}.~Lindstr{\"o}m.
\newblock Common montane birds are declining in northern {E}urope.
\newblock \emph{Journal of Avian Biology}, 45\penalty0 (1):\penalty0 3--14, 2014.

\bibitem[Leito et~al.(2016)Leito, Leivits, Leivits, Raet, Ward, Ott, Tullus, Rosenvald, Kimmel, and Sepp]{leito2016black}
A.~Leito, M.~Leivits, A.~Leivits, J.~Raet, R.~Ward, I.~Ott, H.~Tullus, R.~Rosenvald, K.~Kimmel, and K.~Sepp.
\newblock Black-headed gull ({L}arus ridibundus {L}.) as a keystone species in the lake bird community in primary forest-mire-lake ecosystem.
\newblock \emph{Baltic Forestry}, 22\penalty0 (1):\penalty0 34--45, 2016.

\bibitem[Lindstr{\"o}m et~al.(2015)Lindstr{\"o}m, Green, Husby, K{\aa}l{\aa}s, and Lehikoinen]{lindstrom2015large}
{\AA}.~Lindstr{\"o}m, M.~Green, M.~Husby, J.~A. K{\aa}l{\aa}s, and A.~Lehikoinen.
\newblock Large-scale monitoring of waders on their boreal and arctic breeding grounds in northern {E}urope.
\newblock \emph{Ardea}, 103\penalty0 (1):\penalty0 3--15, 2015.

\bibitem[Lindstr{\"o}m et~al.(2019)Lindstr{\"o}m, Green, Husby, K{\aa}l{\aa}s, Lehikoinen, Stjernman, et~al.]{lindstrom2019population}
{\AA}.~Lindstr{\"o}m, M.~Green, M.~Husby, J.~A. K{\aa}l{\aa}s, A.~Lehikoinen, M.~Stjernman, et~al.
\newblock Population trends of waders on their boreal and arctic breeding grounds in northern {E}urope.
\newblock \emph{Wader Study}, 126\penalty0 (3):\penalty0 200--216, 2019.

\bibitem[Lovette and Hochachka(2006)]{lovette2006simultaneous}
I.~J. Lovette and W.~M. Hochachka.
\newblock Simultaneous effects of phylogenetic niche conservatism and competition on avian community structure.
\newblock \emph{Ecology}, 87\penalty0 (sp7):\penalty0 S14--S28, 2006.

\bibitem[Merkle et~al.(2019)Merkle, Furr, and Rabe-Hesketh]{merkle2019bayesian}
E.~C. Merkle, D.~Furr, and S.~Rabe-Hesketh.
\newblock Bayesian comparison of latent variable models: Conditional versus marginal likelihoods.
\newblock \emph{Psychometrika}, 84:\penalty0 802--829, 2019.

\bibitem[Monti et~al.(2015)Monti, Duriez, Arnal, Dominici, Sforzi, Fusani, Gr{\'e}millet, and Montgelard]{monti2015being}
F.~Monti, O.~Duriez, V.~Arnal, J.-M. Dominici, A.~Sforzi, L.~Fusani, D.~Gr{\'e}millet, and C.~Montgelard.
\newblock Being cosmopolitan: evolutionary history and phylogeography of a specialized raptor, the {Osprey Pandion haliaetus}.
\newblock \emph{BMC Evolutionary Biology}, 15:\penalty0 1--15, 2015.

\bibitem[Niku et~al.(2019)Niku, Hui, Taskinen, and Warton]{niku2019gllvm}
J.~Niku, F.~K. Hui, S.~Taskinen, and D.~I. Warton.
\newblock \texttt{gllvm}: Fast analysis of multivariate abundance data with generalized linear latent variable models in \texttt{R}.
\newblock \emph{Methods in Ecology and Evolution}, 10\penalty0 (12):\penalty0 2173--2182, 2019.

\bibitem[Ovaskainen et~al.(2017)Ovaskainen, Tikhonov, Norberg, Guillaume~Blanchet, Duan, Dunson, Roslin, and Abrego]{ovaskainen2017make}
O.~Ovaskainen, G.~Tikhonov, A.~Norberg, F.~Guillaume~Blanchet, L.~Duan, D.~Dunson, T.~Roslin, and N.~Abrego.
\newblock How to make more out of community data? {A} conceptual framework and its implementation as models and software.
\newblock \emph{Ecology Letters}, 20\penalty0 (5):\penalty0 561--576, 2017.

\bibitem[Piirainen et~al.(2023)Piirainen, Lehikoinen, Husby, K{\aa}l{\aa}s, Lindstr{\"o}m, and Ovaskainen]{piirainen2023species}
S.~Piirainen, A.~Lehikoinen, M.~Husby, J.~A. K{\aa}l{\aa}s, {\AA}.~Lindstr{\"o}m, and O.~Ovaskainen.
\newblock Species distributions models may predict accurately future distributions but poorly how distributions change: {A} critical perspective on model validation.
\newblock \emph{Diversity and Distributions}, 29\penalty0 (5):\penalty0 654--665, 2023.

\bibitem[Poworoznek et~al.(2021)Poworoznek, Anceschi, Ferrari, and Dunson]{poworoznek2021efficiently}
E.~Poworoznek, N.~Anceschi, F.~Ferrari, and D.~Dunson.
\newblock Efficiently resolving rotational ambiguity in {Bayesian} matrix sampling with matching.
\newblock \emph{arXiv preprint arXiv:2107.13783}, 2021.

\bibitem[Preston(1948)]{preston1948commonness}
F.~W. Preston.
\newblock The commonness, and rarity, of species.
\newblock \emph{Ecology}, 29\penalty0 (3):\penalty0 254--283, 1948.

\bibitem[Royle et~al.(2005)Royle, Nichols, and K{\'e}ry]{royle2005modelling}
J.~A. Royle, J.~D. Nichols, and M.~K{\'e}ry.
\newblock Modelling occurrence and abundance of species when detection is imperfect.
\newblock \emph{Oikos}, 110\penalty0 (2):\penalty0 353--359, 2005.

\bibitem[Schein et~al.(2014)Schein, Paisley, Blei, and Wallach]{schein2014inferring}
A.~Schein, J.~Paisley, D.~M. Blei, and H.~Wallach.
\newblock Inferring polyadic events with {Poisson} tensor factorization.
\newblock In \emph{Proceedings of the NIPS 2014 Workshop on” Networks: From Graphs to Rich Data}, 2014.

\bibitem[Sk{\'o}rka et~al.(2014)Sk{\'o}rka, Martyka, W{\'o}jcik, and Lenda]{skorka2014invasive}
P.~Sk{\'o}rka, R.~Martyka, J.~D. W{\'o}jcik, and M.~Lenda.
\newblock An invasive gull displaces native waterbirds to breeding habitats more exposed to native predators.
\newblock \emph{Population Ecology}, 56:\penalty0 359--374, 2014.

\bibitem[Skrondal and Rabe-Hesketh(2004)]{skrondal2004generalized}
A.~Skrondal and S.~Rabe-Hesketh.
\newblock \emph{Generalized latent variable modeling: {Multilevel}, longitudinal, and structural equation models}.
\newblock Chapman and Hall/CRC, 2004.

\bibitem[Steenweg et~al.(2019)Steenweg, Hebblewhite, Whittington, and McKelvey]{steenweg2019species}
R.~Steenweg, M.~Hebblewhite, J.~Whittington, and K.~McKelvey.
\newblock Species-specific differences in detection and occupancy probabilities help drive ability to detect trends in occupancy.
\newblock \emph{Ecosphere}, 10\penalty0 (4):\penalty0 e02639, 2019.

\bibitem[Stratton et~al.(2024)Stratton, Hoegh, Rodhouse, Green, Banner, and Irvine]{stratton2024clustering}
C.~Stratton, A.~Hoegh, T.~J. Rodhouse, J.~L. Green, K.~M. Banner, and K.~M. Irvine.
\newblock Clustering and unconstrained ordination with {D}irichlet process mixture models.
\newblock \emph{Methods in Ecology and Evolution}, 15\penalty0 (9):\penalty0 1720--1732, 2024.

\bibitem[Ter~Braak and Prentice(1988)]{ter1988theory}
C.~J. Ter~Braak and I.~C. Prentice.
\newblock A theory of gradient analysis.
\newblock \emph{Advances in Ecological Research}, 18:\penalty0 271--317, 1988.

\bibitem[Tilman(1982)]{tilman1982resource}
D.~Tilman.
\newblock \emph{{Resource Competition and Community Structure}}.
\newblock Number~17. Princeton University Press, 1982.

\bibitem[van~der Veen et~al.(2023)van~der Veen, Hui, Hovstad, and O'Hara]{van2023concurrent}
B.~van~der Veen, F.~K. Hui, K.~A. Hovstad, and R.~B. O'Hara.
\newblock Concurrent ordination: {Simultaneous} unconstrained and constrained latent variable modelling.
\newblock \emph{Methods in Ecology and Evolution}, 14\penalty0 (2):\penalty0 683--695, 2023.

\bibitem[Vehtari et~al.(2017)Vehtari, Gelman, and Gabry]{vehtari2017practical}
A.~Vehtari, A.~Gelman, and J.~Gabry.
\newblock Practical {Bayesian} model evaluation using leave-one-out cross-validation and {WAIC}.
\newblock \emph{Statistics and Computing}, 27:\penalty0 1413--1432, 2017.

\bibitem[Wang and Zhang(2012)]{wang2012nonnegative}
Y.-X. Wang and Y.-J. Zhang.
\newblock Nonnegative matrix factorization: {A} comprehensive review.
\newblock \emph{IEEE Transactions on Knowledge and Data Engineering}, 25\penalty0 (6):\penalty0 1336--1353, 2012.

\bibitem[Warton et~al.(2015)Warton, Blanchet, O’Hara, Ovaskainen, Taskinen, Walker, and Hui]{warton2015so}
D.~I. Warton, F.~G. Blanchet, R.~B. O’Hara, O.~Ovaskainen, S.~Taskinen, S.~C. Walker, and F.~K. Hui.
\newblock So many variables: joint modeling in community ecology.
\newblock \emph{Trends in Ecology \& Evolution}, 30\penalty0 (12):\penalty0 766--779, 2015.

\bibitem[Watterson(1974)]{watterson1974models}
G.~A. Watterson.
\newblock Models for the logarithmic species abundance distributions.
\newblock \emph{Theoretical Population Biology}, 6\penalty0 (2):\penalty0 217--250, 1974.

\bibitem[Weiher et~al.(2011)Weiher, Freund, Bunton, Stefanski, Lee, and Bentivenga]{weiher2011advances}
E.~Weiher, D.~Freund, T.~Bunton, A.~Stefanski, T.~Lee, and S.~Bentivenga.
\newblock Advances, challenges and a developing synthesis of ecological community assembly theory.
\newblock \emph{Philosophical Transactions of the Royal Society B: Biological Sciences}, 366\penalty0 (1576):\penalty0 2403--2413, 2011.

\bibitem[Yee and Hastie(2003)]{yee2003reduced}
T.~W. Yee and T.~J. Hastie.
\newblock Reduced-rank vector generalized linear models.
\newblock \emph{Statistical Modelling}, 3\penalty0 (1):\penalty0 15--41, 2003.

\bibitem[Yoo and Choi(2009)]{yoo2009probabilistic}
J.~Yoo and S.~Choi.
\newblock Probabilistic matrix tri-factorization.
\newblock In \emph{2009 IEEE International Conference on Acoustics, Speech and Signal Processing}, pages 1553--1556. IEEE, 2009.

\bibitem[Zhou(2015)]{pmlr-v38-zhou15a}
M.~Zhou.
\newblock {Infinite Edge Partition Models for Overlapping Community Detection and Link Prediction}.
\newblock In G.~Lebanon and S.~V.~N. Vishwanathan, editors, \emph{Proceedings of the Eighteenth International Conference on Artificial Intelligence and Statistics}, volume~38 of \emph{Proceedings of Machine Learning Research}, pages 1135--1143, San Diego, California, USA, 09--12 May 2015. PMLR.
\newblock URL \url{https://proceedings.mlr.press/v38/zhou15a.html}.

\bibitem[Zhou et~al.(2024)Zhou, Gu, and Dunson]{zhou2024bayesian}
Y.~Zhou, Y.~Gu, and D.~B. Dunson.
\newblock {Bayesian Deep Generative Models for Replicated Networks with Multiscale Overlapping Clusters}.
\newblock \emph{arXiv preprint arXiv:2405.20936}, 2024.

\bibitem[Zito and Miller(2024)]{zito2024compressive}
A.~Zito and J.~W. Miller.
\newblock Compressive {B}ayesian non-negative matrix factorization for mutational signatures analysis.
\newblock \emph{arXiv preprint arXiv:2404.10974}, 2024.

\end{thebibliography}

\appendix
\renewcommand\thefigure{\thesection.\arabic{figure}}    
\setcounter{figure}{0}  

\section*{Acknowledgments}
We thank Ossi Nokelainen for refining and verifying our interpretations on bird distributions and their relationships to bird resource use. This research was partially supported by the National Institutes of Health (grant ID R01ES035625), by the European Research Council under the European Union’s Horizon 2020 research and innovation programme (grant agreement No 856506), the Office of Naval Research (N00014-21-1-2510; N00014-24- 1-2626), and the National Science Foundation (IIS-2426762).

\section*{Code and Data Availability}
Code and data are available at \url{https://github.com/braden-scherting/barcode}

\newpage

\section{Supplementary Material}\label{sec:extras}
\subsection{Supporting Results}
\subsubsection*{Results using different factorization ranks}
\begin{figure}
\begin{center}
\includegraphics[width=0.95\textwidth]{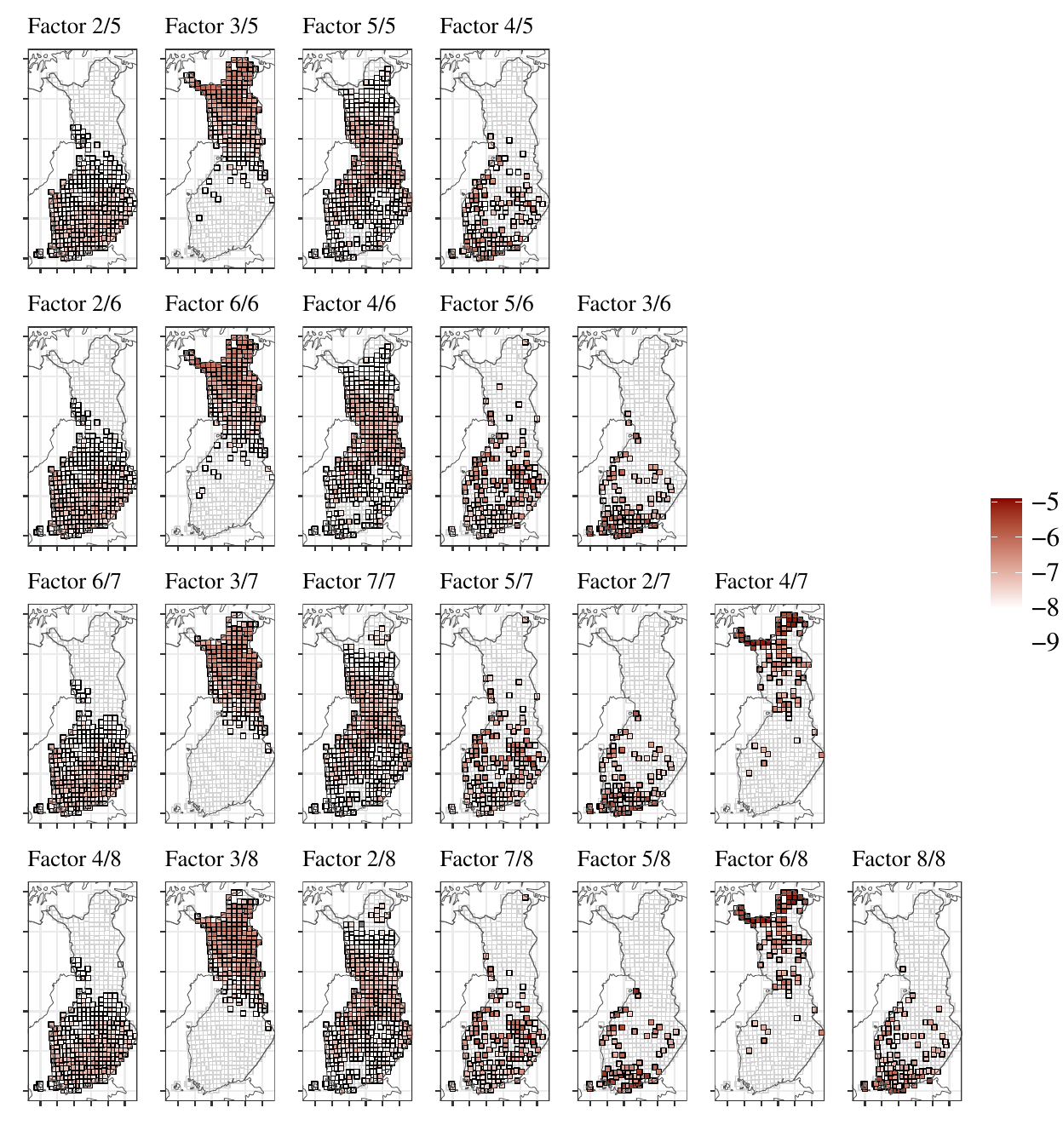}
\end{center}
\caption{\label{fig:maps5678} Factor presence and strength overlaid on the map of Finland obtained using different factorization ranks ($k =5,6,7,8$). Factors were reordered to showcase the correspondence across model fits. Sampling sites are shown as tiles. Black borders indicate the presence of each factor, and coloration indicates log-transformed factor strength. For sites that are sampled multiple times, border and color values represent averages over samples within each site. }
\end{figure}
The main analysis was conducted using results from a rank-7 factorization. As noted in Section 4, this choice balances interpretability against model fit. Figure A.1 displays maps analogous to Figure 1 obtained by fitting with 5, 6, 7, and 8 factors. Again, all model fits include a constant reference factor. Factors are reordered to align similar factors across model fits. The rank-7 maps are the same as those presented in Section 4. Although variation in model fit and the precise interpretation of factors is expected to vary with the number of factors used, we find that many of the same patterns are identified across models. The main analysis identified three factors that broadly partition Finland by latitude and climate and three factors that identify more specialized habitats. Based on the additional maps, all factorizations include factors that align closely with the climatic partitioning. Higher ranks refine the decomposition by adding novel factors, splitting existing factors, or both. The transition from five factors to six seemingly sharpens the distinction between urban and agricultural sites; the Arctic/wetland factor emerges when seven factors are used; the eighth factor is correlated with urban and agricultural factors. Overall, these additional model fits indicate that the qualitative identities of dominant drivers are not an artifact of our chosen factorization rank.

\subsubsection*{Posterior vs empirical marginal expectation}
\begin{figure}[H]
\begin{center}
\includegraphics[width=0.95\textwidth]{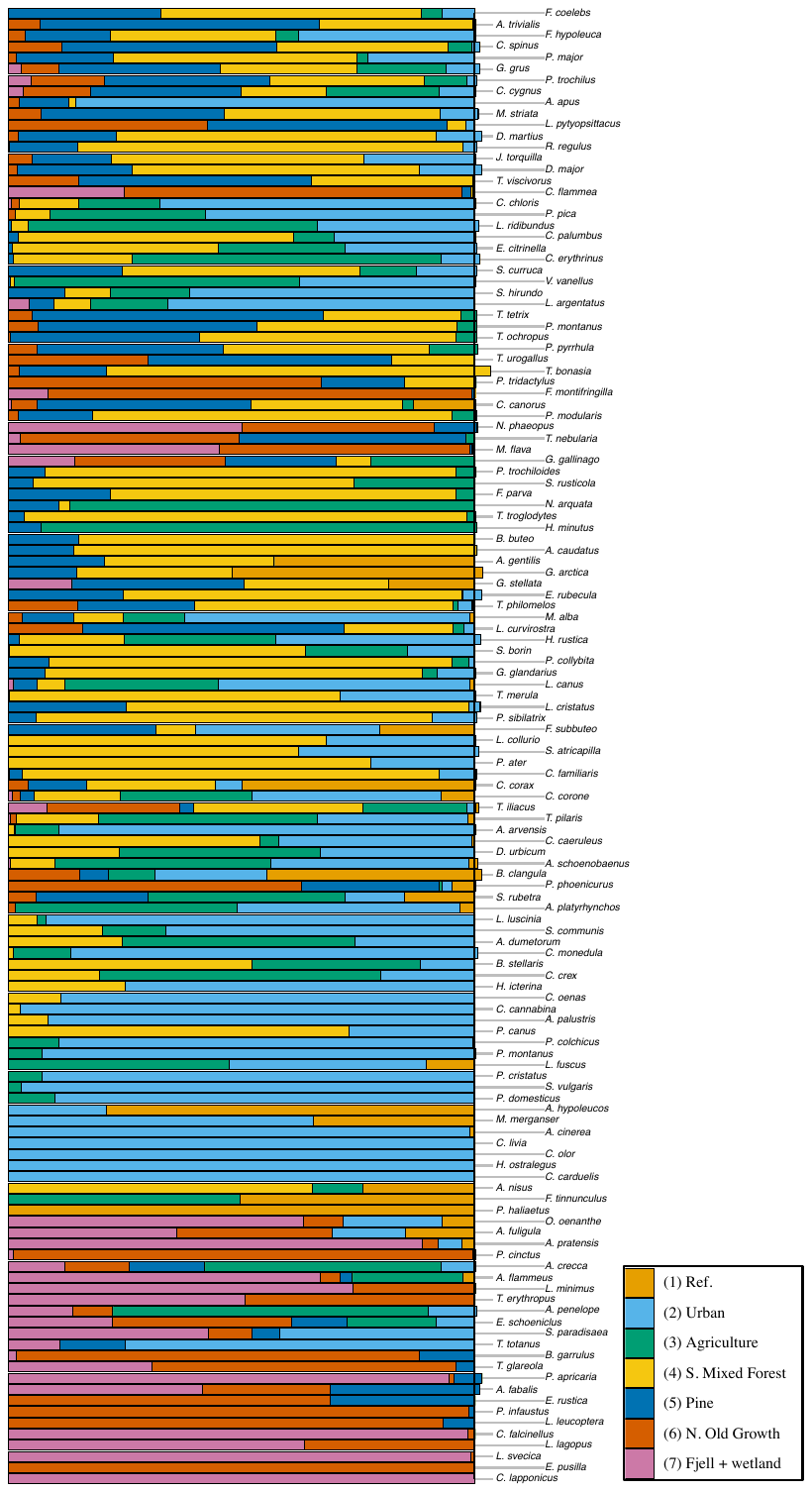}
\end{center}
\caption{\label{fig:margMean} Species profiles clustered by preference and segmented by preference strength. For each species, the width of each bar indicates the posterior marginal expectation relative to the empirical marginal expectation. The vertical black line denotes equality between posterior and empirical marginal expectation.}
\end{figure}

Figure \ref{fig:margMean} serves as a model checking and inferential tool in analogy to Figure 4. For all species, the posterior marginal expectation closely aligns with the average number of individuals observed across samples. As a basic posterior predictive check, this signals good model fit. The same inferences about habitat specialization and species clusters can be drawn here, but this result lacks detail about the extent to which the suite of learned factors explain observed variation.

\subsubsection*{Marginal correlation}
Here, we compare the sample marginal (interspecific) correlation to the marginal correlation under the posterior. The model-based marginal covariance is defined in Section 3.3. In addition to model checking, this comparison serves to certify that barcode captures prevail- ing patterns of dependence between species, including negative dependencies. Importantly, these measures of dependence are not “residual” (correlation between species abundances that cannot be attributed to observed abiotic factors). The model-based correlation esti- mate largely agrees with the sample correlation, but the estimated correlations generally have greater magnitude. Many of the same parameters control the marginal mean, variance, and correlation, which limits the model’s flexibility. If inference on marginal correlations were central to the analysis, we may consider alternative modeling approaches.

\begin{figure}
\begin{center}
\includegraphics[width=0.85\textwidth]{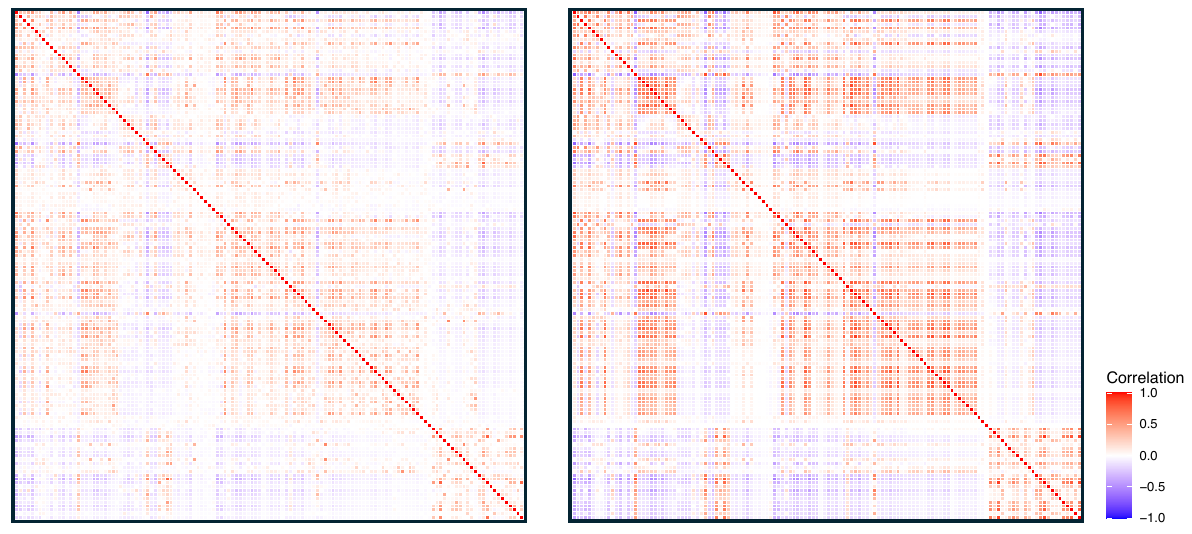}
\end{center}
\caption{\label{fig:corrCompare} Comparison of sample correlation (left) and posterior expected correlation (right) between species. Species names are omitted; the order of species is consistent with the order given in other figures.}
\end{figure}

\subsubsection*{Sampling details}
Here, we present model fitting details for the results presented in Section 4. After initialization (see Section A.4), we generate 50,000 samples from each of four chains. For each chain, the first 25,000 samples are discarded as burn-in, and a further 25,000 samples are drawn and thinned by 10 for memory considerations. This yields a total of 10,000 posterior samples. As noted in Supplement A.4, chains are not initialized at dispersed locations. However, the potential scale reduction factor diagnostic statistic \citep{gelman1992inference} remains useful to confirm that all chains are exploring the same region.

\begin{table}
    \centering
    \begin{tabular}{|c|l|l|l|l|l|}
    \hline 
         & $\ell$ & $C\circ \Phi^*$ & $S\circ \Gamma$ & $B$ &$\Xi$  \\
         \hline 
         PSRF & 1.032 &(1.000, 1.012)&(1.000, 1.042)&(1.001, 1.028)&(1.000, 1.021) \\ 
         \hline
    \end{tabular}
    \caption{Potential scale reduction factors (PSRF) for log-likelihood ($\ell$) and 2.5 and 97.5 elementwise PSRF quantile for model parameters. *Approximately 15\% of factor switches cil are equal to zero with probability $< 1/10000$ \textit{a posteriori}, so PSRF is undefined; the reported range reflects convergence diagnostic statistics for those $(i, l)$ which have nonzero posterior expected value.}
    \label{tab:placeholder}
\end{table}

\subsection{Simulations}

\begin{figure}
\begin{center}
\includegraphics[width=1\textwidth]{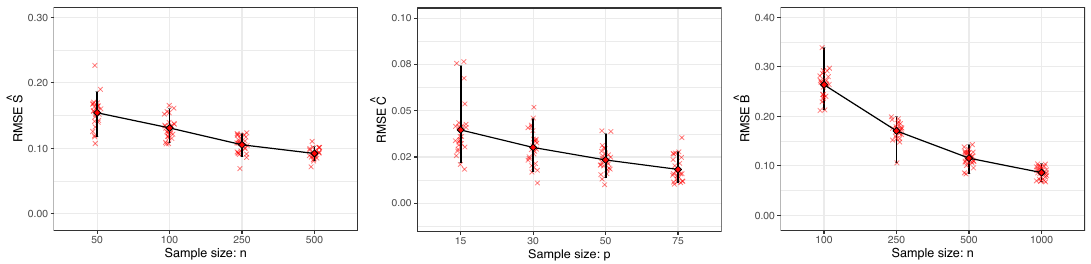}
\end{center}
\caption{\label{fig:threeRMSE} Simulation results for $S$ (left), $C$ (middle), and $B$ (right) are displayed in the top row. Each ($\times$) denotes the average error under the posterior, averaged across all elements of $S$ ($C, B$, respectively). Medians, 5\% quantiles, and 95\% quantiles are indicated by the red diamond, lower bar, and upper bar respectively.}
\end{figure}

\begin{figure}
\begin{center}
\includegraphics[width=1\textwidth]{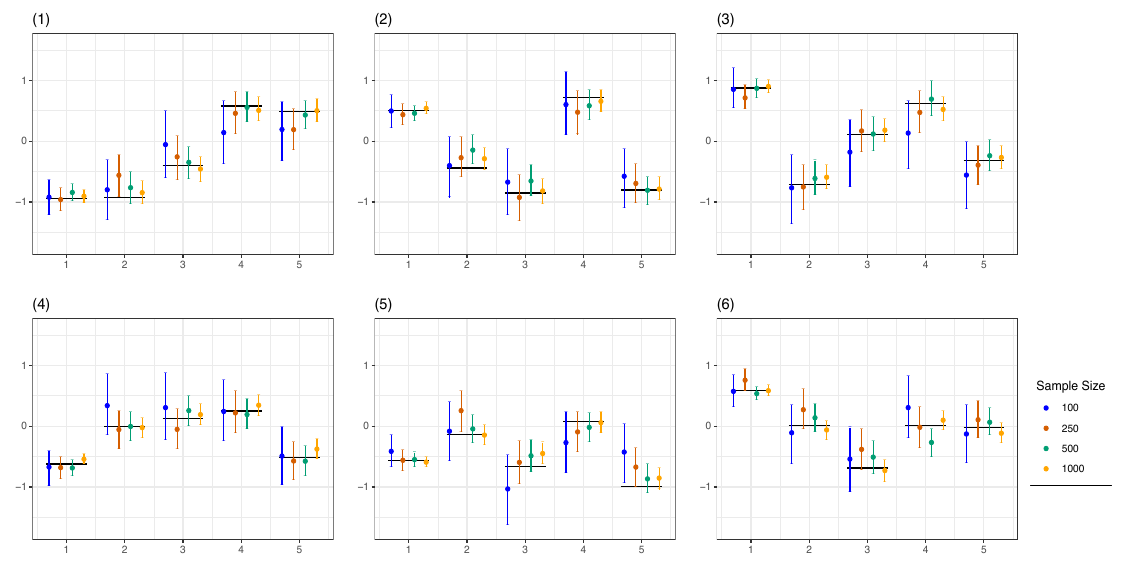}
\end{center}
\caption{\label{fig:betaRecoAll} Detailed estimates of B from one of the 25 replicate simulations. For each replicate, five covariates are simulated (x-axis) for each of the six non-constant factors ((1)–(6)).}
\end{figure}

We investigated the performance of barcode through simulation to certify the model’s ability to recover sparse latent variables and latent regression coefficients. These parameters are central to interpretable inference. We first evaluate recovery of the binary latent variables $C$ and $S$. Varying $n$ affects the effective sample size for estimating $s_j$ and varying $p$ affects the effective sample size for estimating $c_i$. Thus, to study $s_j$ , we let $n = (50, 100, 500, 1000)$ with $p = 50$, and to study $c_i$, we let $p = (15, 30, 50, 75)$ and $n = 500$. Under these conditions, we
generate $C$ and $S$, ensuring that no rows of $C$ or $S$ are all zero (that is, no rows or columns of $Y$ are deterministically zero). We then sample $\Phi$ and $\Gamma$ from Ga($1,\frac{1}{3}$) and Ga($1,\frac{1}{5}$) respectively, and draw $Y$ element-wise from appropriate conditional Poisson distributions. The distributions used to simulate $\Phi$ and $\Gamma$  are chosen to have large variances to induce very large and very small non-zero values in $M$. For each $(n,p)$, we repeat this process 25 times. The model is fit without covariates, which may be regarded as model-based unconstrained ordination. We do not observe label switching but we permute columns of $\hat{C}$ and $\hat{S}$ to make direct comparisons. Figure \ref{fig:threeRMSE} shows the estimation error for $S$ and $C$. We additionally simulate under the concurrent \texttt{barcode} model (Equation 3), varying $n = (100,250,500,1000)$, to confirm that latent covariate effects on factor presence are recoverable. For each n, 25 data sets are simulated. The results are shown in Figure \ref{fig:threeRMSE} c. Posterior credible intervals reliably cover the true value and narrow as n increases. Figure \ref{fig:betaRecoAll} displays estimates of $B$ for one of the 25 data sets. The results of these simulations are somewhat unsurprising: the correctly specified model reliably recovers data-generating parameters. However, this reveals an attractive property of discrete latent variables: $S$ and $C$ are identifiable up to permutation. This is consistent with recent theoretical developments on identifiability in a wide range of latent class models \citep{gu2023bayesian, lee2024new}.

\subsection{Selecting the number of factors $L$}

The factorization rank $L$ can be chosen to facilitate interpretation, set based on information criteria and fit to data, or estimated as part of an expanded joint model. WAIC is a good choice for comparing Bayesian latent factors models with different factorization ranks, because it can be readily computed, does not directly depend on the number of model parameters, and approximates out-of-sample prediction accuracy; see \cite{merkle2019bayesian} for details and discussion of its use in different practical settings, and \cite{vehtari2017practical} for computational considerations. For a prediction-based approach tailored to nonnegative matrix factorization that employs regularization and cross-validation directly, see \cite{lal2021novo}. Analogous regularization could be applied through a prior. For both WAIC and cross-validation, the model must be refit for each level of regularization and/or factorization rank. This can be avoided by using a hyperprior that controls shrinkage and the number of factors directly.
Estimating the number of latent factors in Gaussian factor models using shrinkage priors has received considerable attention \citep{carvalho2008high, legramanti2020bayesian, fruhwirth2023generalized}. \cite{zito2024compressive} develop an approach tailored to the Poisson factorization model using a compressive hyperprior that shrinks the factor loadings matrix column-wise. Adopting this prior for $\Gamma$ would be natural, as they also normalize factor scores. The matrix $S$ can also indirectly induce partial sparsity in redundant factors. To achieve exact column-wise sparsity, $S$ could be thresholded based on corresponding values of $\Gamma$ - e.g., if $\gamma_{jl}<\epsilon$, $s_{jl} = 0$. The threshold could also be applied to entire columns based on the estimated relevance weights under their compressive prior framework. Alternatively, factor-specific preference probabilities $\{\psi_l\}$ could be introduced and appropriate shrinkage priors assigned to induce sparsity directly through $S$, an approach related to spike-and-slab priors.

\subsection{Computation details}\label{sec:CompDetails}

Our product formulation of factors $c_{il}\phi_{il}$ and loadings $s_{jl}\gamma_{jl}$ naturally facilitates Gibbs sampling, but a pseudo-prior analogous to those first introduced for variable selection by \cite{dellaportas2002bayesian} is useful for tuning mixing properties. Our pseudo-prior is
$$
(\gamma_{jl}\mid  s_{jl}=0, -)\sim \text{Ga}(1, \tau_0),
$$
This choice does not affect the posterior. To further encourage exploration of the discrete latent space, we jointly update blocks of binary switches. At each iteration and for each $i = 1,\dots n$ (resp. $j = 1,\dots,p$), factor indices $\{1,\dots,L\}$ are randomly partitioned into $b$ blocks $\{l\}_b$ of size $n_b$. Switches within each block $\{c_{il} : l \in \{l\}_b\}$ ($\{s_{jl} : l \in \{l\}_b\}$) are jointly updated by drawing a configuration from the set of $2^{n_b}$ configurations. This is made simpler by the fact that posterior probabilities of candidate configurations are trivial to compute, and some configurations can be automatically rejected. A candidate $\hat{\bm{c}}_i$ ($\hat{\bm{s}}_j$ induces $\{\hat{\mu}_{i1},\dots,\hat{\mu}_{ip}\}$ ($\{\hat{\mu}_{1j},\dots,\hat{\mu}_{nj}\}$). If any $y_{ij} > 0$ such that $\hat{\mu}_{ij}=0$, the candidate can be excluded from consideration.

Poisson factorization sometimes yields posteriors with multiple modes that are not solely attributable to rotational ambiguity, owing to the non-convexity of the joint likelihood. This is a computational challenge which, if not addressed, complicates inference. Empirically, we find that several non-equivalent configurations of latent variables separated by regions of low probability can sometimes reasonably explain the data, particularly in sparse settings. However, the configurations are generally distinct (i.e., significantly different in log-posterior), and one typically dominates. Our objective in this study is to find the scientifically meaningful configuration or configurations of factors that best explain abundance and characterizing corresponding uncertainty, rather than exhaustively exploring and characterizing uncertainty across all modes. The main obstacle to doing so is the potential for MCMC chains to become stuck in different local modes. This is a well-known property of nonnegative matrix factorization and Bayesian variants, which has led scholars to develop and study initialization procedures \citep{langville2006initializations}. \cite{zito2024compressive} ran multiple chains and used samples only from the chain with the highest log-posterior. We opt to initialize $\Phi$ and $\Gamma$ using the fast optimization-based Poisson factorization of \cite{cortes2018fast}. We rescale factorization matrices so that $\sum_i \phi_{il}=1$, and set $C=\bm{1}_n\bm{1}_L^\top$ and $S=\bm{1}_p\bm{1}_L^\top$. Following initialization, burn-in sampling is still performed. Although the sampler typically navigates away from the warm starting point, we find this approach to perform better than random initialization or other intialization schemes (e.g., \cite{gaujoux2010flexible}) in terms of post-burn-in log-posterior and sampling time.

\end{document}